\documentclass[prb,twocolumn]{revtex4}
\pdfoutput=1
\usepackage[applemac]{inputenc} 
\usepackage{graphicx} 
\usepackage{epsfig} 
\usepackage{epstopdf}
\usepackage{amssymb}
\usepackage{axodraw}

\newcommand{\bea}{\begin{eqnarray}}
\newcommand{\eea}{\end{eqnarray}}
\newcommand{\beq}{\begin{equation}}
\newcommand{\eeq}{\end{equation}}
\newcommand{\benu}{\begin{enumerate}}
\newcommand{\enu}{\end{enumerate}}

\newcommand{\si}{\sigma}

\newcommand{\ham}{\mathcal{H}}

\begin{document}
\title{He$^3$ bi-layers as a simple example of de-confinement}
\date{\today}
\author{A. Benlagra, C. P\'epin}
\affiliation{Institut de Physique Th\'eorique,
CEA, IPhT, F-91191 Gif-sur-Yvette, France
CNRS, URA 2306, F-91191 Gif-sur-Yvette, France \\
}
\begin{abstract}
We consider the recent experiments on He$^3$ bi-layers\cite{saunders}, showing 
evidence for a quantum critical point (QCP) at which the first layer localizes.
Using the Anderson lattice in two dimensions with the addition of a small dispersion of 
the f-fermion, we modelize the  system of adsorbed He$^3$ layers. The first layer represents the
f-fermions at the brink of localization while the second layer behaves as a free Fermi sea. 
We study the quantum critical regime of this system, evaluate the effective mass in the Fermi liquid phase and
the coherence temperature and give a fit of the experiments and interpret its main features.  Our model can serve as well as a predictive tool
used for better determination of the experimental parameters.
\end{abstract}

\leavevmode
\maketitle

\section{Introduction}
 In the last fifteen years, an increasing body of experiemental results has revealed remarkable properties in heavy fermions, close to a zero temperature phase transition\cite{stewart,lohneysen,review-piers}.  The standard laws governing the behavior of  metallic conductors at very low temperature appeared to be violated in heavy  fermions.  Proximity to a QCP was  early invoked, to explain the experiments \cite{review-piers,rosch} but so far, this wide body of observations remains a mystery and a challenging open problem.
 
\begin{figure}[h]
\includegraphics[width=5cm]{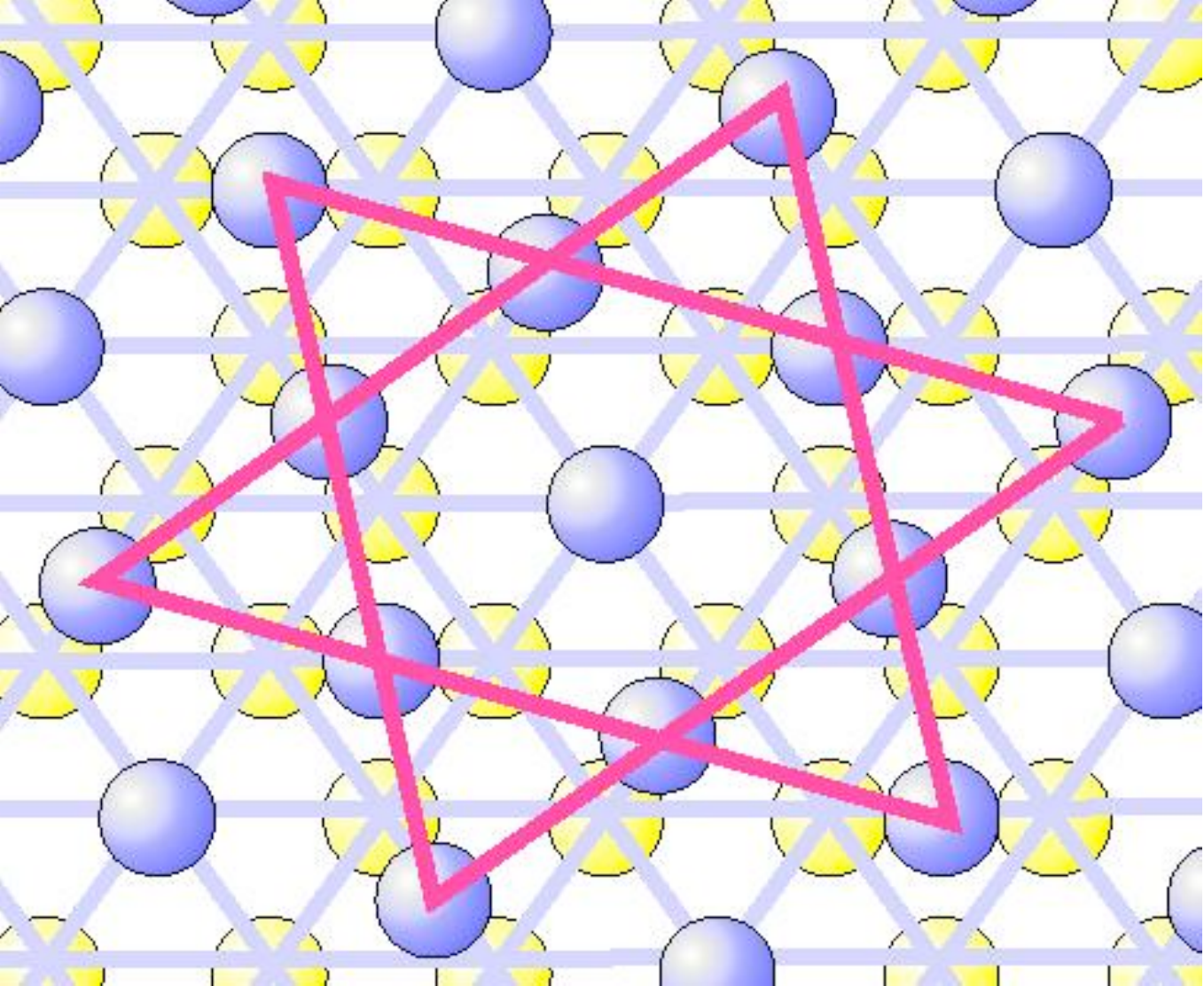}
\caption{$^3$He solid layer on top of the triangular lattice of the substrate} \label{superlattice}
\end{figure}

 Recently, a new experimental set-up was explored, showing signs of quantum criticality of the same nature as for heavy fermions\cite{saunders}, but in a rather different system. It consists of  two layers of He$^3$ fermions adsorbed on two layers of He$^4$; those themselves adsorbed on a graphite substrate. The history of He$^3$ films adsorbed on graphite is quite rich \cite{godfrin,greywall,godfrin2, godfrin3}.   A first layer of He$^3$ has been adsorbed on graphite in two typical situations; on top of a compressed He$^4$ solid of density $11.2  nm^{-2}$ and on top of a deuterium layer  of density $9.1 nm^{-2}$. In both cases, a solidification of the  top He$^3$ layer is observed  at a ratio of densities $N_{1}/N_{sub}= 4/7 $. This ``magic'' number corresponds to  a half filled super-lattice of unit cell $\sqrt{7} \times \sqrt{7}$ (see Fig. \ref{superlattice}), formed on top of the triangular substrate lattice.
        Specific heat measurements show that the effective mass increases by a factor of  ten in the approach of the transition. The magnetic structure of the localized phase has been extensively studied.  It is believed to be a spin liquid induced by ring-exchange\cite{roger1}. The  precise determination of this spin liquid phase, and particularly  whether it is massless or massive, and whether it has some ferromagnetic component is still under debate\cite{gregoire}. Then a second and third  He$^3$ layers were adsorbed. The originality of the experiment \cite{saunders}  is that it is the first time that, when the second layer arrives  at promotion, the first layer has not yet solidified. Hence there is a regime in coverage where the two first  layers hybridize while layer one sits on the brink of localization. 

Experimental details can be found in Ref. \cite{saunders}.  We give here a rapid summary of the main findings of this work.
The second layer arrives at promotion at a total coverage of  $N= 6.3 nm^{-2}$.  From $6.3 $ to $9.2 nm^{-2}$, a carachteristic temperature $T_{0}$ is extracted, from the specific heat measurements, below which the fluid bilayer has Fermi liquid properties with an enhanced quasiparticle mass. Above $T_{0}$, a Curie law is observed, as if the first layer de-confines  from the heavy Fermi liquid and behaves as a localized spin while the second one behaves as a Fermi liquid. It is quite difficult however, to separate quantitatively the contribution of each layer in the heavy Fermi liquid phase. This characteristic temperature seems to vanish at a coverage $N_{crit} = 9.95 nm^{-2}$, the so-called "critical coverage" by the experimentalists, with a power law
\beq
T_{coh} \sim \delta^{1.8} \  ,\eeq
where $\delta  = | N_{crit}  - N| /N_{crit}$. 

The effective mass is shown to increase by  a factor of 18 at $N=9.0 nm^{-2}$  and seems to diverge at $N_{crit}$ with a power law \beq m/m^* \sim  \delta  \ . \eeq 

Beyond $N_c$, the first layer is fully localized at all temperatures investigated. 
However, NMR studies show that at $N_{I}= 9.2 nm^{-2}$ the magnetization starts to grow in a rather abrupt manner.  It is not excluded that a first order ferromagnetic transition occurs for $N \geq N_{I}$ but an experimental evidence for it is still not conclusive.
The localized phase is believed to be a spin liquid ; a small ``bump'' in the heat capacity marks the onset of the spin liquid  parameter. Experimentally it is evaluated to  be of the order of  $J \sim 7 \ m K $.   Last, an activation gap is extracted from the heat capacity measurements. It decreases with increasing coverages and there are indications that it vanishes at  a coverage lower than $N_{crit}$.

  In this  paper we give the details of the calculations whose results have been announced in a previous Letter\cite{adel}. We apply the theory of the Kondo breakdown, previously introduced for the study  of QCP in heavy fermions\cite{us,cath,cathlong} to the system of He$^3$ bi-layers.
   The formalism is identical to the one developed in \cite{cathlong}. We use the Anderson lattice model with  the addition of a dispersion of the f-fermions, to describe the system of He$^3$ bi-layers.  The first layer, in the brink of localization,  forms the lattice of  f-fermions.  When the first layer localizes, the lattice is half-filled by construction. Strong hard core repulsion is taken into account by a short range Coulomb repulsion $U$, with $U \sim  20 K$, in agreement with the early studies of bulk He$^3$ \cite{vollhardt}. The top layer is modeled as a free Fermi gas. Hybridization between the two layers consists of hopping processes from layer one to layers two and vice versa \cite{heritier}. 
   
    The paper is organized as follows. In section \ref{sec:model} we present the Anderson lattice model and derive the slave-boson effective Lagrangian. Section \ref{sec:parameters} is devoted to the evaluation of the bare parameters' dependence in coverage. This is necessary if we want to confront our theory to the experimental data. We present in section \ref{sec:MFT} the mean-field approximation. We show the presence of a QCP at $T=0$ corresponding to the Mott localisation of He$^3$ first layer's fermions. In particular, a peculiar behavior of the effective hybridization explains the apparent occurence of two QCPs in the experimental data. We then study the fluctuations in section \ref{sec:fluc} discussing the critical regime and computing the effective mass and the coherence temperature in an intermediate energy regime corresponding to a dynamical exponent $z =3$. We conclude in section \ref{sec:conclusions} with our main result and give a criticism of our work.
    Some technical details are presented in the appendices. Appendix A shows the calculation of the integrals used at the mean-field approximation. In Appendix B, we give the details of the evaluation of the fermionic contribution to the corrections of scaling of the holon mass and discuss the stability of the QCP. Finally, in appendix C, we derive an expression of the free energy starting from the Luttinger-Ward formula.
    
    \section{The model}
    \label{sec:model}Our starting point is the Anderson lattice model

\bea \ham   & = & \sum_{
\langle i, j \rangle, \sigma} \left [  {\tilde f}^\dagger_{i \sigma}  \left (t^0_{ij} + (E_0 - \mu) \delta_{ij} \right ) {\tilde f}_{j \sigma}  \right  . \nonumber \\
& + & \left . 
 c^\dagger_{i \sigma} \left ( t_{ij} - \mu \delta_{ij} \right )  c_{j \sigma} \right ]   
+   V   \sum_{i \sigma}     \left   (  {\tilde f}^\dagger_{i \sigma} c_{i \sigma}  + h.c. \right ) \nonumber \\
&  + &  
\sum_i \left ( U {\tilde n}_{f, i}^2 + U_1 {\tilde n}_{f, i } n_{c, i } \right )  \ , \eea 

Here $\langle i,j \rangle $ refers to nearest neighbour sites created by the Graphite's corrugate potential, $\si$ is a spin index, ${\tilde f}^{\dagger}_{i \si} ({\tilde f}_{i \si})$ are creation (annihilation) operators for the first layer's fermions,  $c^{\dagger}_{i \si} (c_{i \si})$ are creation (annihilation) operators for the second layer's fermions. $t_{ij}= t$ is the c-fermion's hopping, $t^{0}_{ij} = \alpha t $ is the f-fermion's hopping term, V is the hybridization between the two layers, $E_{0}$ is the energy level of the f-fermions and $ \mu $ is the chemical potential. ${\tilde n_{f, i}} = \sum_{\si} {\tilde f}^{\dagger}_{i \si} {\tilde f}_{i \si}$ and  ${ n_{c, i}} = \sum_{\si} {c}^{\dagger}_{i \si} {c}_{i \si}$ are the operators describing the particle number of each layer's fermions. $U$ and $U_{1}$ are respectively the intra- and inter-layer Coulomb repulsion. The model is studied in the limit of very large on-site repulsion $U$. We expect to have a coverage dependant hopping parameter, $t \equiv t(N)$ as well as hybridization  $V(N)$. Furthermore, we have $U1 \ll U$, but we keep the inter-layer interaction term for now.

Super-exchange terms can be generated by a second order expansion in large $U/(\alpha t)$ and $U_{1}/(\alpha t)$. The Hamiltonian is then written
\bea \ham   & = & \sum_{
\langle i, j \rangle, \sigma} \left [  {\tilde f}^\dagger_{i \sigma}  \left (t^0 + (E_0-\mu) \delta_{ij} \right ) {\tilde f}_{j \sigma}  \right  . \nonumber \\
& + & \left . 
 c^\dagger_{i \sigma} \left ( t - \mu \delta_{ij} \right )  c_{j \sigma} \right ]   
+   V   \sum_{i \sigma}     \left   (  {\tilde f}^\dagger_{i \sigma} c_{i \sigma}  + h.c. \right )  \\
&  + &  
\sum_{\langle i,j\rangle} J \,\,  \left( \mathbf{\tilde S}_{f,i}.\mathbf{\tilde S}_{f,j} - {\tilde n_{i}}{\tilde n_{j}}/4 \right ) + J_{1} \mathbf{\tilde S}_{f,i}.\mathbf{ S}_{c,j}  \ , \nonumber \label{hamiltonian} \eea 
where $J= 2(\alpha t)^2/U, J_{1}= 2(\alpha t)^2/U_{1} $ and $\mathbf{\tilde S}_{f}=\sum_{\alpha, \beta}f^{\dagger}_{\alpha} \mathbf{\sigma}_{\alpha \beta} f_{\beta}$ is the spin operator with $\vec \sigma$ the Pauli matrix. RKKY interaction, mediated by the conduction electrons, as well as various ring exchange parameters, studied in \cite{roger2} can be included in the J-term.

 One key approximation of this work is that we consider that at the edge of localization, the f-fermions are half-filled. This means that the f-fermion somehow form their ``own'' lattice as the coverage increases, so that when the localization occurs, we are at half filling.  This approximation is necessary if we want to attribute the observed  increase of the effective mass to strong correlations coming from Mott physics.  However, we don't have a microscopic justification for it; only the coherence of the findings of this approach can justify it. The on-site Coulomb repulsion U is very large ( $ \sim 20 \ K $), leading to strong correlations effects. In the limit $U \rightarrow \infty$, there is a constraint of no double occupancy which we account for using Coleman's slave boson \cite{coleman84}, decoupling the f-fermion's creation operator at each site ``i'' as
\beq
{\tilde f}^{\dagger}_{i \si} \rightarrow   f^{\dagger}_{i \si} b_{i} \label{coleman}
\eeq
where $f^{\dagger}$, the creation operator of the so-called ``spinons'' and $b^{\dagger}$, the one of the holons, obey the local constraint $\sum_{\si}f^{\dagger}_{i \si}f_{i \si} + b^{\dagger}_{i}b_{i} = 1$. Upon the transformation (\ref{coleman}), the slave boson drops of all bi-linear products of fields at the same site.

The constraint is taken into account in a Lagrangian formulation through a Lagrange multiplier $\lambda$.The effective lagrangian is then
\bea \mathcal{L}   & = & \sum_{
\langle i, j \rangle, \sigma} \left [  { f}^\dagger_{i \sigma}  \left ((\partial_{\tau}+ \epsilon_{f,i})\delta_{ij} +  b_{i} t^0 b^{\dagger}_{j}  \right ) {\tilde f}_{j \sigma}  \right  . \nonumber \\
& + & \left . 
 c^\dagger_{i \sigma} \left ( (\partial_{\tau}- \mu) \delta_{ij} + t  \right )  c_{j \sigma} \right ]  \nonumber \\
& + &
 \sum_{i} b^{\dagger}_{i}(\partial_{\tau} + \lambda_{i})b_{i} - \lambda_{i} +  V   \sum_{i \sigma}     \left   (  { f}^\dagger_{i \sigma}b_{i} c_{i \sigma}  + h.c. \right ) \nonumber \\
&  + &  
\sum_{\langle i,j\rangle} J \,\, \left( \mathbf{S}_{f,i}.\mathbf{S}_{f,j} - {n_{i}}{ n_{j}}/4 \right ) + J_{1} \mathbf{ S}_{f,i}.\mathbf{ S}_{c,j} \ , \eea 
where $\epsilon_{f,i} = E_{0} - \mu + \lambda_{i}$ is the renormalized f-band's chemical potential.
\newline

 The short range magnetic interaction and the induced Kondo interaction
 are decoupled using Hubbard-Stratanovich transformations : $J \mathbf{S}_{f,i}.\mathbf{S}_{f, j} \rightarrow \phi_{i,j}f^\dagger_{i\sigma}f_{j\sigma}- |\phi_{ij}|^2/J$ and $J_{1} \mathbf{S}_{f,i}.\mathbf{S}_{c, j} \rightarrow \sigma_{i}f^\dagger_{i\sigma}c_{i\sigma}- |\sigma_{i}|^2/{J_{1}}$.
\newline

The Lagrangian becomes now

\bea \mathcal{L}   & = & \sum_{
\langle i, j \rangle, \sigma} \left [  { f}^\dagger_{i \sigma}  \left ((\partial_{\tau}+ \epsilon_{f,i} )\delta_{ij} +  b_{i} t^0 b^{\dagger}_{j}  + \phi_{ij} \right ) {\tilde f}_{j \sigma}  \right  . \nonumber \\
& + & \left . 
 c^\dagger_{i \sigma} \left ( (\partial_{\tau}- \mu) \delta_{ij} + t  \right )  c_{j \sigma} \right ]  \nonumber \\
& + &
 \sum_{i} b^{\dagger}_{i}(\partial_{\tau} + \lambda_{i})b_{i}  +    \sum_{i \sigma}     \left   (  { f}^\dagger_{i \sigma}(V b_{i}+ \sigma_{i}) c_{i \sigma}  + h.c. \right ) \nonumber \\
& - &
\sum_{i}( \lambda_{i} +  |\sigma_{i}|^2/(J_{1}) - \sum_{\langle i,j \rangle}|\phi_{ij}|^2/J    \eea 

We assume that $\phi_{ij}$ condenses in a uniform spin liquid phase, i.e. $\langle \phi_{ij} \rangle = \phi_{0}$. It renormalizes the dispersion of the spinon band and ensures the breakdown of the Kondo effect\cite{us}. It is shown to stay roughly constant through the phase diagram $\phi_{0}=\beta t \approx J$\cite{cathlong}. $\sigma_{i}$ merely renormalizes the effective hybridization $V b_{i}$.

    \section{The parameters}
    \label{sec:parameters}

    Before going further, we need to evaluate the dependance of the bare parameters in coverage in order to fit the experimental data.

\begin{figure}[h]
\includegraphics[width=5cm]{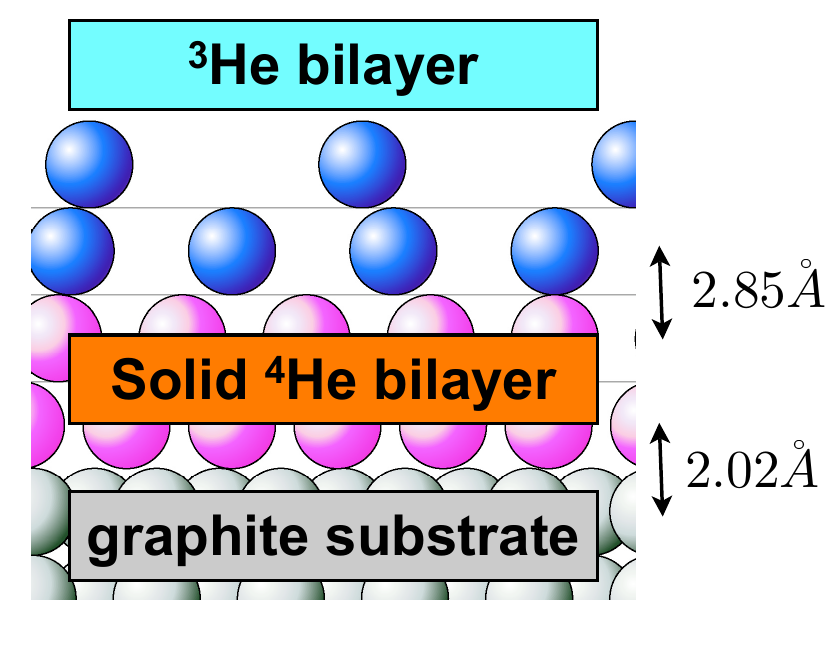}
\caption{$^3$He solid layers on top of the triangular lattice of the substrate; we show here the various heights of the layers one compared to the other.} \label{layers}
\end{figure}

 The height of the layers is taken from the study by Roger \textit{et al.}\cite{roger2}: the first $^4$He layer's height is $\approx 2.02 \, \AA$ while the others' one is $\approx 2.85 \,\AA$ (see Fig. \ref{layers}). From the experiment \cite{saunders}, the density of He$^4$ layers is  $9.2 \, nm^{-2}$ while the one of the first He$^3$ layer is $N_{1} = 6.3 \,nm^{-2}$.
    The total coverage is defined as 
 \bea
 N  & = &  N_{c} + N_{f} \ , \nonumber  \\
 N_{f} & = & N_{1}(1-n_{b}) \ , \eea  where  $N_f$ and $N_c$ are respectively the coverage  ( in $nm^{-2}$)  of the first and second layers and $n_b$ is the number of holons per site.
 At the transition, we have 
 \beq
 N_f /N_{1} = 1 \eeq  which accounts for the fact that at the transition, the f-fermions are in a 1/2 filled lattice.  This means that the number of holons $n_b$ vanishes at the QCP. Away from the QCP, the number of holons is allowed to fluctuate freely and its value is determined self-consistently.
 
 The parameter $J$ is extracted from the experiment \cite{saunders}: $J = 7 mK.$
\newline
\newline

The evaluation of the bandwidth, $D=2t = \pi /m$, for each layer of He$^3$ is based on an analysis in  Pricaupenko and Treiner\cite{pricaupenko} where the kinetic energy of liquid He$^3$ contains a density dependant effective mass : 
\beq \frac{\hbar^2}{2m^*} = \frac{\hbar^2}{2m} \left ( 1 - \frac{\bar \rho}{ \rho_{c}}\right )^2,\label{density}\eeq where $\bar \rho = 3/(4 h_{c}) n(\AA^{-2})$ is the average density inside a sphere of radius $h_{c}=2.63 \AA$ and $\rho_{c} = 0.04 \AA^{-3}$. 

We have then 
\bea
D_{f} = D \,(1 - 0.07\, N_{f})^2 \nonumber \\
D_{c} = D \,(1 - 0.07\, (N - N_{f}))^2 \, , \nonumber
\label{hopping}
\eea
where D is the bandwidth of $^3$He in the bulk. 

At half filling,   $N_{f} = N_{1}$, the mean kinetic energy  $E_{kin, f}$ equals the bandwidth $D_{f}$. We have 
\begin{eqnarray}
E_{kin, f}& =& \frac{\hbar^2}{2m^{*}_{f}N_{1}} \int^{k_{F}}_{0}\frac{d^2\mathbf{k}}{(2\pi)^2} k^2,\nonumber\\
& = & \frac{\hbar^2 \pi^3}{16 m^{*}_{f}} N_{1}
\end{eqnarray}
where $k_{F}=\pi/\sqrt{N_{1}}$ ($\sqrt{N_{1}}/2$ is the average radius of a particle in the first layer).

We find then at half-filling
\bea
E_{c,f}& \approx & 0.52 \, K \nonumber \\
D_{f} &=& 0.30 D \nonumber \\
D_{c} &=& 0.62  D \nonumber \ , 
\eea

thus, $D \approx 1.73 K$, $D_{f} \approx 0.57 \, K$ and $D_{c} \approx 1.18 \, K$ which gives a value $\alpha = 0.54$ for the ratio between the bandwidths. This value is relatively high compared with the typical values for rare earth compounds for which $\alpha \equiv 0.1$. 
\newline

A word has to be said at this stage : we have considered the spherical dispersion of the free fermions
$$\epsilon_{\mathbf{k}} = \frac{k^2}{2m}-\frac{k_F^2}{2m}$$
for which the density of states (DOS), defined by $\frac{d^2k}{4\pi^2}=\rho(\epsilon)d\epsilon$, is constant
$$\rho(\epsilon)=\frac{m}{2\pi}.$$
However, as emphasized in the introduction, the first layer solidifies into a triangular lattice. For a triangular lattice tight-banding band structure, the dispersion is given by
$$\epsilon_{\mathbf{k}} = -2 t \left ( \cos{(k_x)} + 2 \cos{(k_x/2)}\cos{(\sqrt{3}k_y/2)} \right ).$$ 

The Fermi surface for fermions in a triangular lattice is no longer circular at each filling, but we can consider that these deviations are benin in the range of coverage studied in our case, in particular very close to the QCP.

 \begin{figure}[ht]
 \begin{tabular}{cc}
  \includegraphics[width=3.75cm]{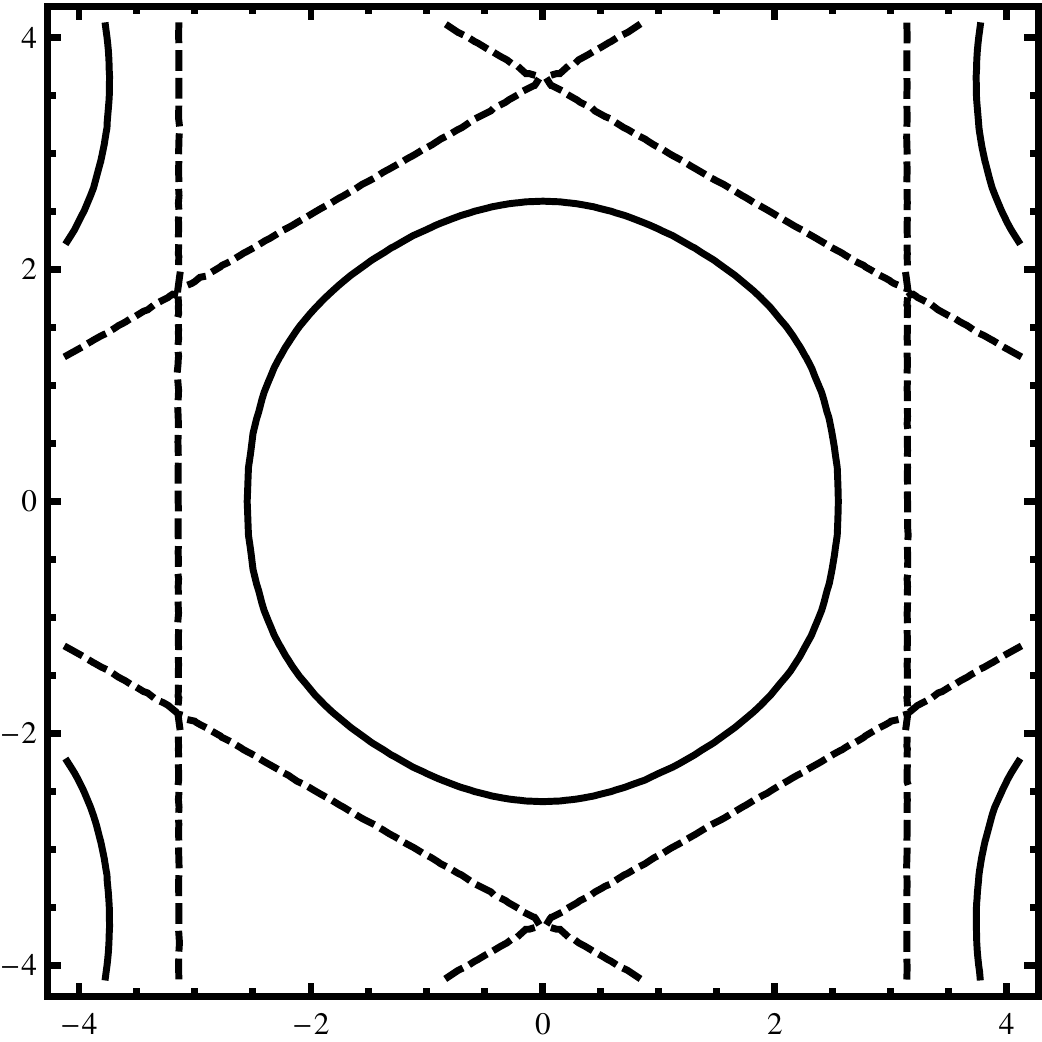} &  \includegraphics[width=3.75cm]{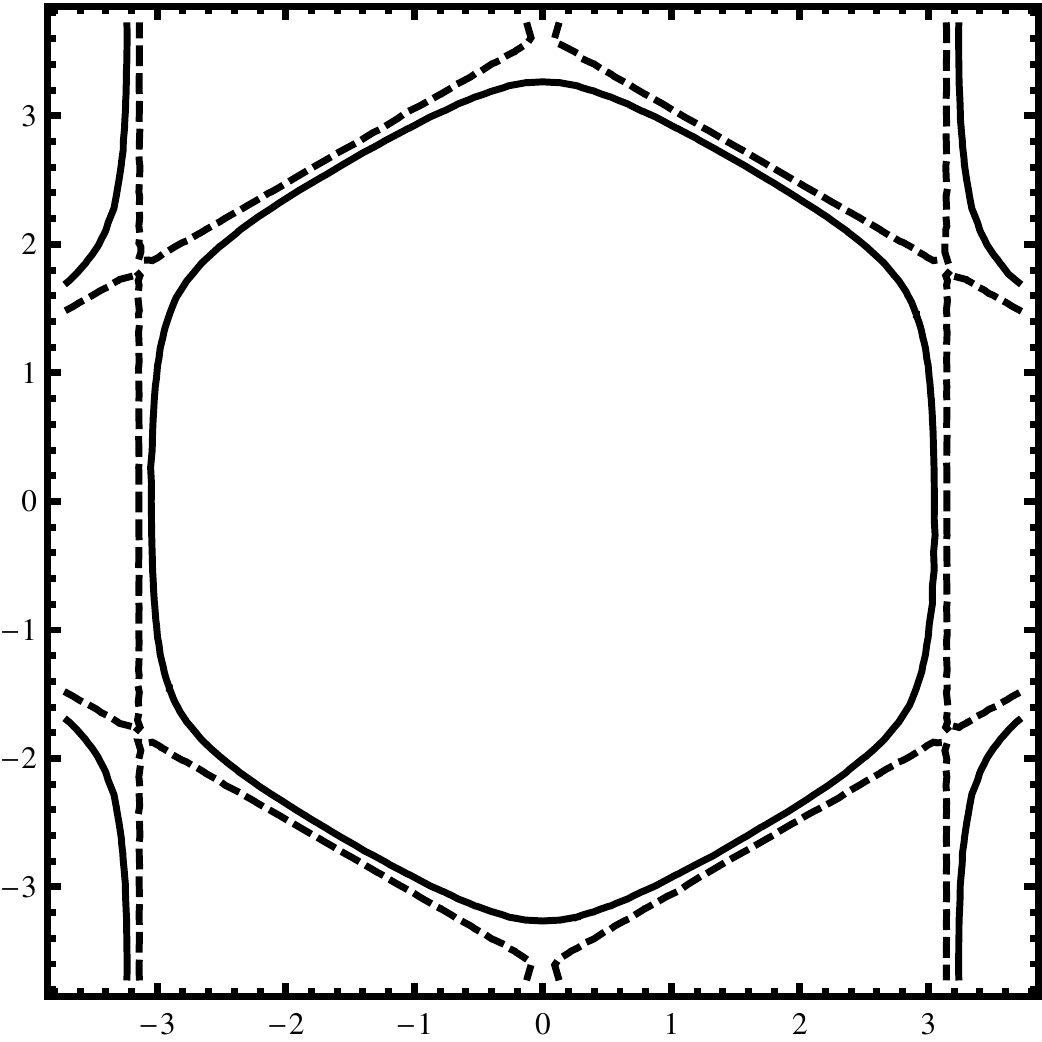} \\
  (a) & (b)
\end{tabular}
\caption{The Fermi surface at unit coverage : (a) $\delta = 0.10$ and (b) $\delta = 0.15$. We see that the Fermi surface in the former is still circular while it experiences, for the second one, small deviation from the circular case. } \label{FS}
\end{figure} 

Fig. (\ref{FS}) shows the Fermi surface of the f-fermions at two different coverages : (a) $\delta = 0.10$  for which $D_f \approx 0.52 K, \epsilon_f \approx 0.13 K$ and (b) $\delta = 0.15$ for which $D_f \approx 0.60 K, \epsilon_f \approx 0.54 K$. In the latter case, the Fermi surface deviates around the circular Fermi surface for free fermions.

\begin{figure}[h]
\includegraphics[width=8.5cm]{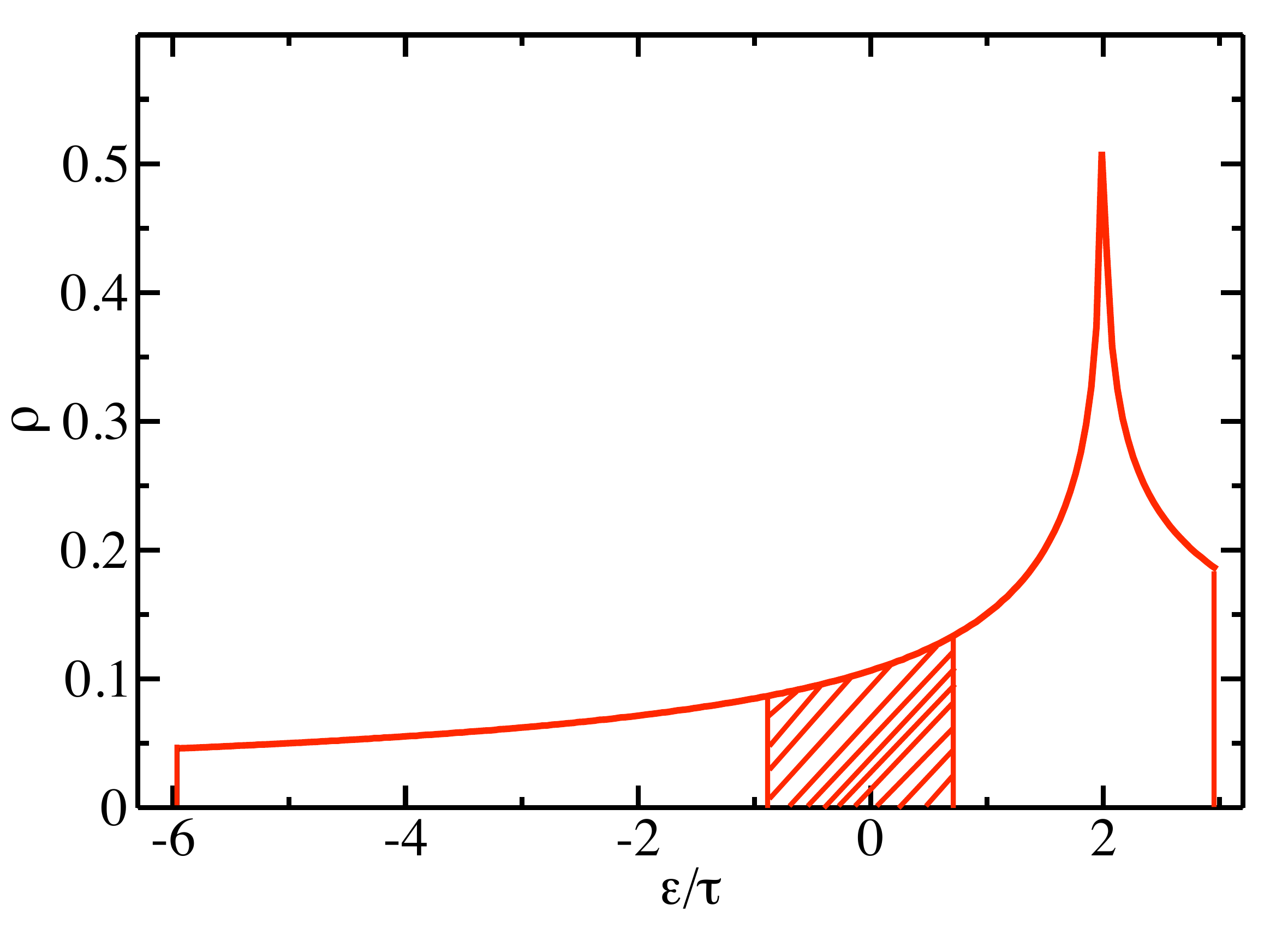}
\caption{Density of states for a triangular lattice tight-banding structure. Characteristic energy scales in our model lie within the hatched region.} \label{dos}
\end{figure}

The approximation of constant DOS can still hold and this can be seen indeed by considering the DOS profile for the triangular lattice case shown in Fig. \ref{dos}.
The hatched region marks the energy scales of our model, and we see that we are far from the Van Hove singularity, and we can approximate the DOS by a constant one.
\newline
\newline

The chemical potential $\mu$ is defined by the filling of the second layer
$$\frac{N_{c}}{6.3}=\int_{D_{c}}^{\mu}\rho_{0}d\epsilon$$
We get directly
\beq \mu = D_{c}(\frac{2N_{c}}{6.3}-1)
\eeq
\newline

    $E_{0}$ is identified as the difference between the potential energies of the two layers \cite{tasaki}. Each layer experiences two kinds of interaction : 
    \begin{itemize}
    \item Van der Waals interaction with the grafoil substrate 
  \beq
  \label{vdv} V_{s}(z)=(4 C^{3}_{3}/{27 D^2})\,1/z^9 - C_{3}/z^3 \ , \eeq where $D = 192 \, K$ is the well depth of the potential and $C_{3}= 2092 \, K \AA^3$ is the Van der Waals constant\cite{pricaupenko}, and 
  
  \item the Bernardes-Lennard Jones interaction between two He particles 
  \beq \label{lj} V_{LJ}(z) = 4 \epsilon \, ((\sigma/r)^{12}-(\sigma/r)^6) \ , \eeq with $\epsilon = 10.2 \, K$ and $\sigma = 2.56 \, \AA$ is the hard core radius \cite{roger2}. Thus, for the layer $L_{i}$, the potential energy writes 

\beq E_{f,c} = V_{s}(z_{i}) + v_i \ , \eeq with 
\beq v_i= \pi \sum_{j} \rho_{j} \int r dr V_{LJ}(r) \ , \eeq where  $\rho_j $  is the density of each layer and ``j''  is the layer's index.

the chemical potential $E_0$ now reads
\beq
E_0 = E_f - E_c \  . \eeq
 
\end{itemize} 

 We denote (see Fig. \ref{layers}) $r_1$, $r_2$, $r_3$ and $r_4$  respectively the distances  of the first, second, third and fourth layers to the graphite center.  We have
 \beq
 \begin{array}{cccc}
 z_1= 2.2 \AA \ , & z_2 = 5.03 \AA \ , & z_3 = 7.9 \AA \ ,  & z_4 = 10. 57 \AA \ . \end{array} \eeq Applying (\ref{vdv}) we get
 $V_s (z_3) = -4.21 \ K $ and $ V_s (z_4) = -1.77  \ K $. These orders of  magnitude are quite big compared to the typical scale of a few mK  for this system. Its order of magnitude is in accordance with \cite{roger2}.
 
 We turn now to the Lennard-Jones potential. We sum up (\ref{lj}) for all two body  interaction  in all the layers. We get for the first layer  or ``f''-fermions
 \bea v_f  & = &   \pi \left [ \rho_1 \int_{r_m^1}^\infty r dr V_{LJ}(r) + \rho_2 \int_{r_m^2}^\infty r dr V_{LJ}(r) \right . \nonumber \\
  & + &\left .  \rho_3 \int_{r_m^3}^\infty r dr V_{LJ}(r)  + \rho_4 \int_{r_m^4}^\infty r dr V_{LJ}(r) \right ] \ ,  \eea with
  \[ \begin{array}{ll}
  \rho_1= 0.092  \ ,& \; \; \; \rho_2 = 0.092 \ , \\ r_m^1= 5.7 \ , & \; \; \;r_m^2 = 2.85 \ ,\end{array} \]
  \[ \begin{array}{ll} \rho_3 = 10^{-2} N_1 (1-n_b) \ , & \; \; \; \rho_4 = 10 ^{-2}[ N- N_1 ( 1-n_b) ]  \ , \\
   r_m^3= \rho_3^{-1/2} \ , & \;  \;  \; r_m^4= 2.85 \ . \end{array} \]
   For the second layer or ``c'' fermions, we get
   \bea v_c  & = &   \pi \left [ \rho_1 \int_{r_m^1}^\infty r dr V_{LJ}(r) + \rho_2 \int_{r_m^2}^\infty r dr V_{LJ}(r) \right . \nonumber \\
  & + &\left .  \rho_3 \int_{r_m^3}^\infty r dr V_{LJ}(r)  + \rho_4 \int_{r_m^4}^\infty r dr V_{LJ}(r) \right ] \ ,  \eea with
  \[ \begin{array}{ll}
  \rho_1= 0.092  \ ,& \; \; \; \rho_2 = 0.092 \ , \\ r_m^1= 8.55 \ , & \; \; \;r_m^2 = 5.7 \ ,\end{array} \]
  \[ \begin{array}{ll} \rho_3 = 10^{-2} N_1 (1-n_b) \ , & \; \; \; \rho_4 = 10 ^{-2}[ N- N_1 ( 1-n_b) ]  \ , \\
   r_m^3= 2.85 \ , & \;  \;  \; r_m^4= \rho_4^{-1/2} \ . \end{array} \]  The values of $r_m^j$ are now in $\AA$.
   \[  \parbox[c]{8cm}{\includegraphics[width=8cm]{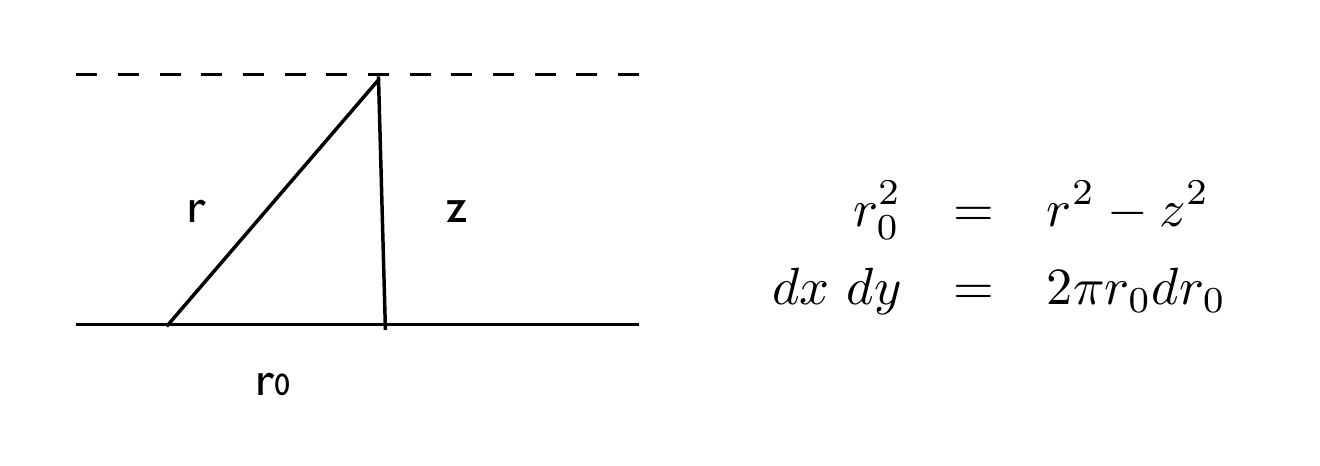}} \]
  We finally  get
  \bea v_f  & = &  -3.87 - 1.08 N - 6.75 n_b + I_3  \\
    \mbox{with}  \;  \; I_3  & = &  -0.196 (1-n_b)^3 [11.2 - 0.3 (1-n_b)^3 ]  \ . \nonumber \eea and
    \bea v_c  & = &  -7.69 + 6.75 n_b + J_4  \\
   \mbox{with}  \;  \; J_4  & = &  - \pi  10 ^{-2}  (N - 6.25 (1-n_b)^3 )  \nonumber \\
   & &   \left [ 0.29 - 3.2 10^{-5}  (N - 6.25 (1-n_b)^3 ) \right ] \  . \nonumber \eea
    $E_0$ now reads
    \bea
E_{0}\equiv & 1.65 -1.071 N -10^{-6}N^2-13.5n_{b} + \nonumber \\
 & -2.25 (1-n_{b})^3 \nonumber \\
 &+0.059(1-n_{b})^6
\eea
\newline


The last parameter, and the most crucial in fact is the hybridization $V$. It is defined as the hopping strength between the two layers. We can have an estimate of V using equation (\ref{density}) to get the same dependence as in (\ref{hopping})
$$V = V_{0} (1 - 0.07\, (N - N_{f})) (1 - 0.07\, N_{f}).$$
Here, $V_{0}$ is proportional to the overlap between the ground state wave functions of the two layers, i.e.
$$t_{12}\approx \delta V \int dz \Phi_{1}(z)\Phi_{2}(z)\, ,$$
where $\Phi^{k}_{i}$ is the ground state wave function of layer $i$ and $\delta V = V_s (z_4)-V_s (z_3)$.

The latter is taken as a Slater determinant of single particle states $\Phi_{i}^{k}$ which writes, assuming translational invariance parallel to the surface \cite{pricaupenko, roger2} :
$$\Phi^{k}_{i}(\mathbf{r}) = \frac{1}{2\pi}\phi^{k}_{i}(z)\exp{[i(k_{x}x+k_{y}y)]}$$
Density functional models show a Lorantzian-like profiles for the density of each layer along the z-direction \cite{pricaupenko} :
$$\rho_{i} \equiv |\phi_{i}(z)|^2 = \frac{b_{i}}{(z-z_{i})^2 + a^2_{i}}$$

From \cite{roger2}, we have :
\bea
\mbox{For $L_{1}$:}& a =1.7 \,\,\,\,\, b=0.115 \nonumber \\
\mbox{For $L_{2}$:}&  a = 3.42 \,\,\,\,\, b=0.32 \nonumber
\eea
\begin{figure}[ht]
\includegraphics[width=5cm]{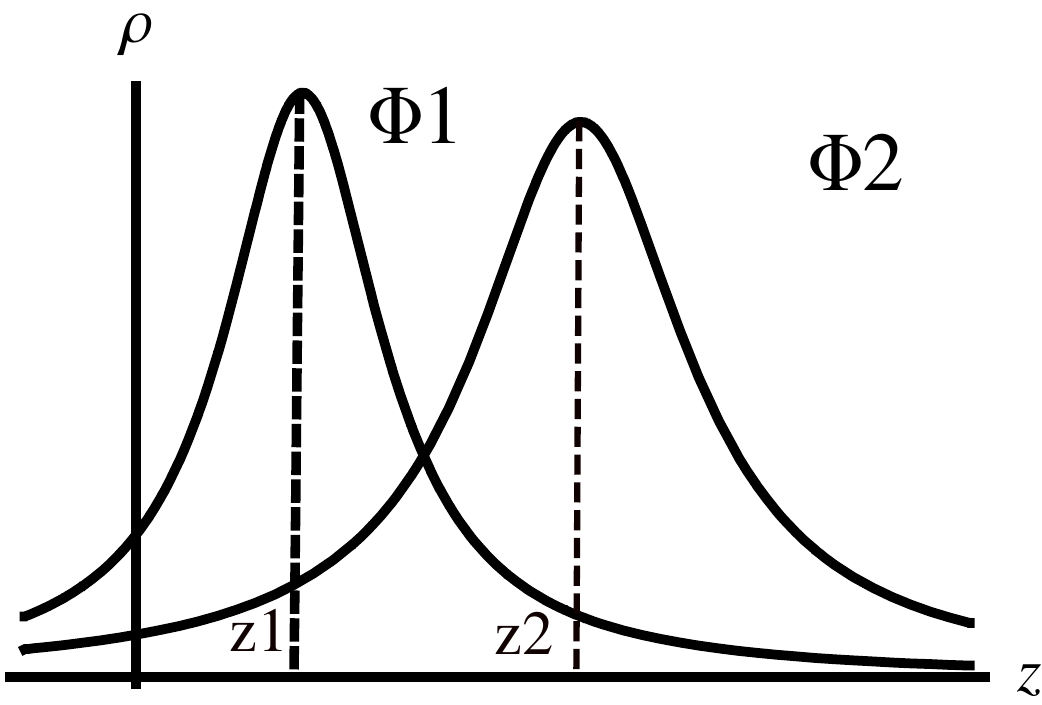}
\caption{Sketch of layers' density profile. The hybridization is estimated from the overlap between the wave functions of the two layers.} \label{hybrid}
\end{figure} 
We find then $V_0 \approx 0.6 K$ consistent with the value obtained in a previous study \cite{tasaki}.
\newline

As said before, the hybridization $V$ is actually a crucial parameter. Indeed, the mean field value for the QCP reads $J/t = \exp{\left [ E_{0}D_{c}/V^2 \right ]}$\cite{us, cathlong}, we see then that any small variation in the dependance of $V$ on coverage has an exponential impact on the position of the QCP. We consider $V$ thus as a fitting parameter that will tune the position of the QCP. 

We have taken $$V = V_{0} (1 - 0.07\, (N - N_{f})) (1 - 0.07\, N_{f}) + V_{1} \delta + V_{2} \delta^2,$$ where $V_{0}$, $V_{1}$ and $V_{2}$ are adjusted to fit the experimental data and $\delta = (N_{crit}-N)/N_{crit}$. We used $V_{0} = 1.55 \, K$, $V_{1}=15.9 \, K$ and $V_{2}=-4.5 \, K$ .
\newline

    \section{Mean-field theory}
    \label{sec:MFT}

At the mean field level, we make a uniform and static approximation for the holon field and the Lagrange multiplier. The free energy writes then 

\bea
F_{MF}& =  -2 T  \sum _{k, \sigma, \omega_{n}, \pm} &  \ln{(-G^{-1}_{\pm}(i \omega_{n}, \mathbf{k}))}  \nonumber \\ & & + \lambda (b^2 - 1) \, 
\label{MFFE}
\eea
 
where $\omega_{n}$ is the fermionic Matsubara frequency and $G^{-1}_{\pm} = i \omega_{n} - E_{\mathbf{k} \pm}$, with

$$E_{\mathbf{k}\pm} = \frac{1}{2}\left [ \epsilon_{\mathbf{k}} + \epsilon^{0}_{\mathbf{k}} \pm \sqrt{(\epsilon_{\mathbf{k}}-\epsilon^{0}_{\mathbf{k}})^2 + 4 V^2 b^2} \right ] \,$$

\begin{figure}[h]
\includegraphics[width=8cm]{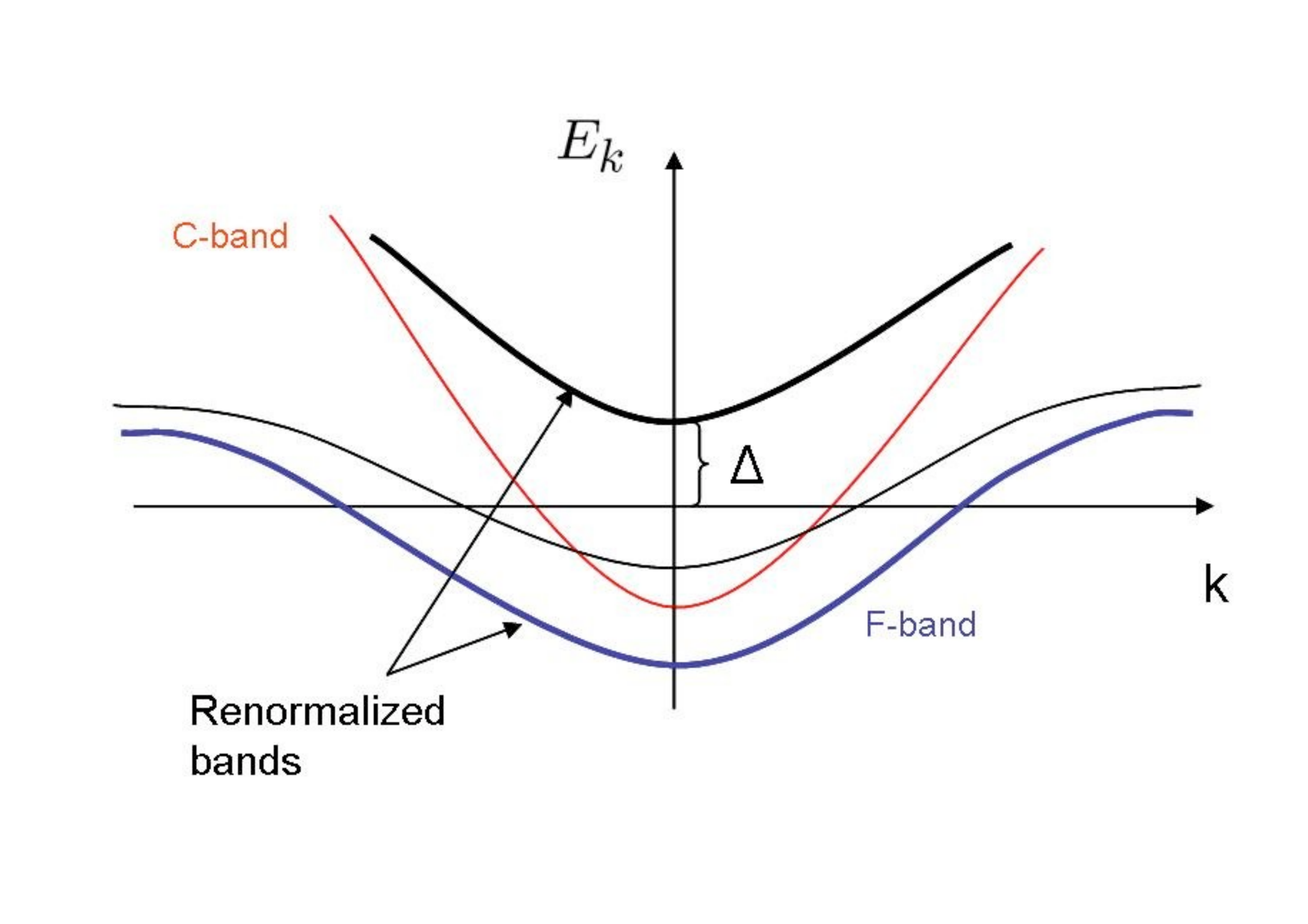}
\caption{Sketch of the different dispersions: the lower band is the dispersion of the spinons and the higher band is the dispersion of the conduction electrons. The Kondo gap $\Delta$ is defined as the difference between the chemical potential and the upper band.} \label{gap}
\end{figure}

In the obove, $e_{\mathbf{k}}$ is the dispersion of the conduction electrons, $\epsilon^0_{\mathbf{k}} = (\alpha b^2 + \beta )\epsilon_{\mathbf{k}} + \epsilon_{f}$ is the spinon dispersion and $E_{\mathbf{k\pm}}$ the dispersion of the renormalized upper (+) and lower (-) bands (See Fig.\ref{gap}) The former derives from the c-fermions with weak f character whereas the latter derives from the f-fermions with weak c character.

Minimizing (\ref{MFFE}) with respect to the holon field $b$ and the Lagrange multiplier $\lambda$, one gets the following mean-field equations

\begin{eqnarray}
T  \sum _{k, \sigma, \omega_{n}} (\alpha \epsilon_{\mathbf{k}} G_{ff} + V G_{fc}) +  \epsilon_{f} - E_{0}  =  0 \nonumber \\
T  \sum_{k, \sigma, \omega_{n}} G_{ff} + b^2  =  1  \ , \label{MFE}\end{eqnarray}

where \bea
G_{ff} & = & \frac{i \omega_{n} - \epsilon^0_{\mathbf{k}}}{(i \omega_{n}-E_{\mathbf{k+}})(i \omega_{n}-E_{\mathbf{k-}})} \,\, , \nonumber \\
G_{fc} &  = & \frac{V b}{(i \omega_{n}-E_{\mathbf{k+}})(i \omega_{n}-E_{\mathbf{k-}})} \, \, , \nonumber \\
 & \equiv & Vb P_{fc}
\eea
These equations are solved in the case of a linearized dispersion bandwidth at zero temperature (T = 0). The summation over $(\mathbf{k}, \omega_{n})$ is performed anatically and is given in Appendix A. The set of resulting equations is then solved numerically.
\newline

\begin{figure}[ht]
\includegraphics[width=8.5cm]{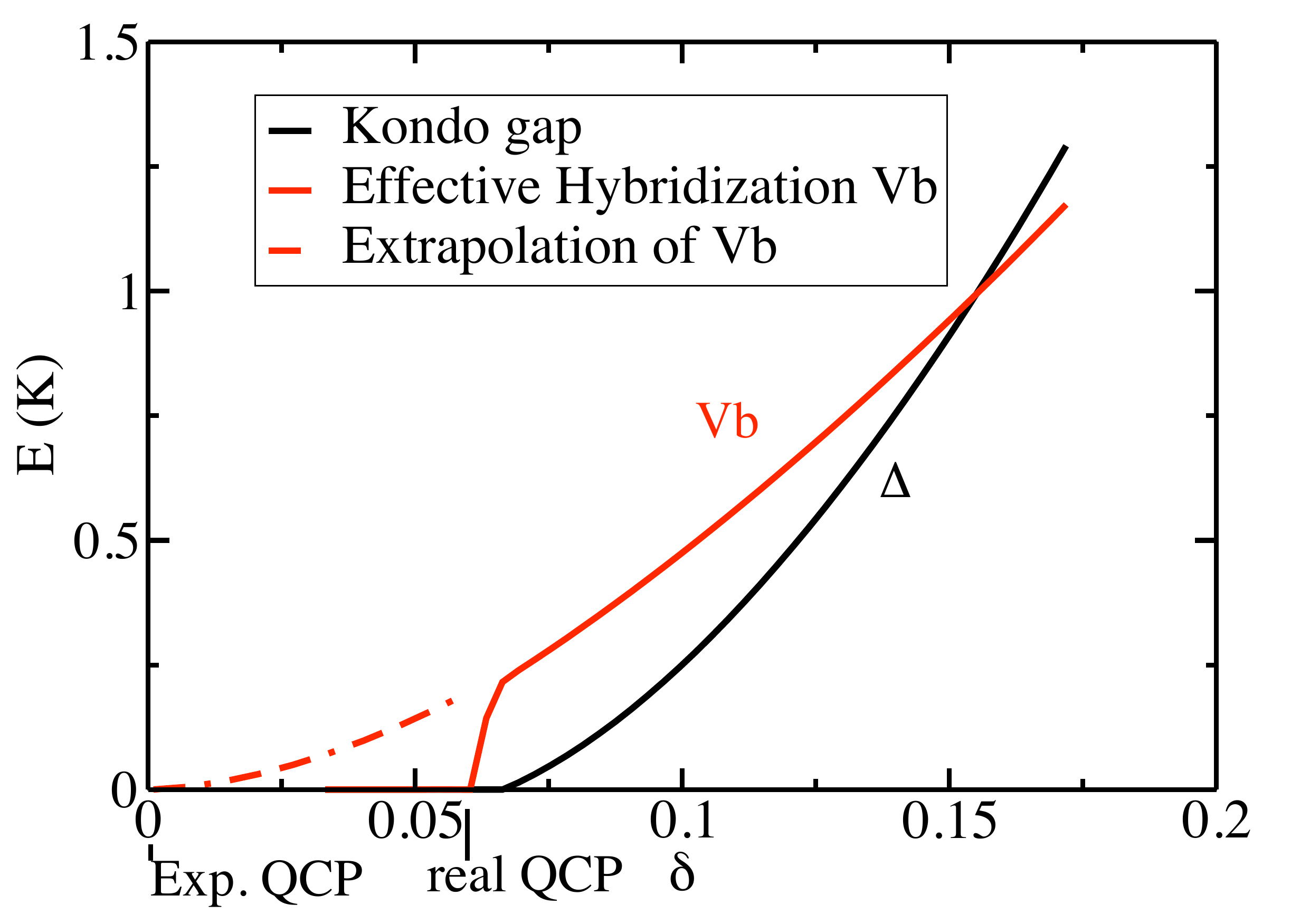}
\caption{Mean-field phase diagram for the Anderson lattice model in D = 2 applied to He$^3$ bilayers \cite{footnote}. Following \cite{saunders} $\delta = 1 - N/{N_{crit}}$ with $N_{crit}=9.95 nm^{-2}$. The effective hybridization (red online) drops suddenly at $\delta \approx 0.063$, indicating the real QCP. The experimental QCP is obtained by extrapolation of Vb to zero (E=0). The Kondo gap $\Delta$ (black online) vanishes before the real QCP \cite{adel}.} \label{MF}
\end{figure}

Fig.\ref{MF} shows the plot of the order parameter, defined as the effective hybridization $Vb$, and the "Kondo gap'' $\Delta$, defined as the energy difference between the chemical potential and the upper band (See Fig. \ref{gap}), as a function of $\delta = 1 - N/{N_{crit}}$.  
\newline

In our model, the Kondo gap is identified with the activation energy observed experimentally in the specific heat.  We have two bands in the  model, one for the spinons and one for the conduction electrons. At very low hybridization,  when the bands  just start to hybridize,  there is no energy difference between the lower and the upper band.  As the hybridization grows, the upper band becomes empty and  an activation gap opens. We see on Fig. \ref{MF}  that the gap closes  at the very vicinity of the QCP.
\newline

The set of mean-field equations shows a QCP where $b \rightarrow 0$, the so-called Kondo Breakdown (KB) QCP, which implies that the spinons experience a Mott transition and their band is half-filled. We observe that $V b$  goes to zero, before the experimentally observed QCP occurs, at a unit coverage $\delta \approx 0.063$.
 This constitutes one main finding of this paper.  The localization occurs before the experimental QCP is reached. Our interpretation is that first, the experimental QCP is evaluated by extrapolating to zero temperature the power laws for the effective mass and the coherence temperature; second, a key feature of the model is that the hybridization is strong, compared to the  other parameters (it is of the order of the bandwidth), hence the falling down of  the order parameter  close to the transition is very abrupt.

This fact is illustrated in Fig. \ref{MF} where we see that the order parameter's behavior has two regimes: it starts to grow very quickely at the QCP then reaches, at the ``elbow'', a regime of strong hybridization. The behavior of the order parameter is governed by the relative strength of the bare hybridization V compared to the other energies of the model. The former is already big at the QCP, $V_c \approx 1.63 K$, thus the slope of the effective hybridization is steep in the hybridized phase. The sharp change corresponds to the emptying of the upper band, the same point at which the opening of the Kondo gap occurs. This point is situated after the real QCP, in the hybridized phase, because when the localization occurs, the f-band is half-filled and the upper band is constrained to sit below the chemical potential and is thus occupied. The vanishing of the Kondo gap before the QCP is observed experimentally if we identify it as the activation gap extracted from the thermodynamic measurements of Neumann \textit{et al.}\cite{saunders}
\newline
 We can make the same construction as the experimentalists, by extrapolating the order parameter in the high energy regime to zero temperature.  We find an additional QCP  that we identify with the ``experimental'' one. This gives an explanation of the mysterious presence of two QCPs  in this system; the magnetization starts to grow at the physical QCP, before the experimental one is reached. Indeed, as soon as the first layer localizes, one expects the static magnetic susceptibility to grow quickly since the spin liquid parameter is small $J \sim 7 mK$.
Note that the  distance in coverage between the two QCPs is in agreement with the experimental data.

\begin{figure}[ht]
\includegraphics[width=7.5cm]{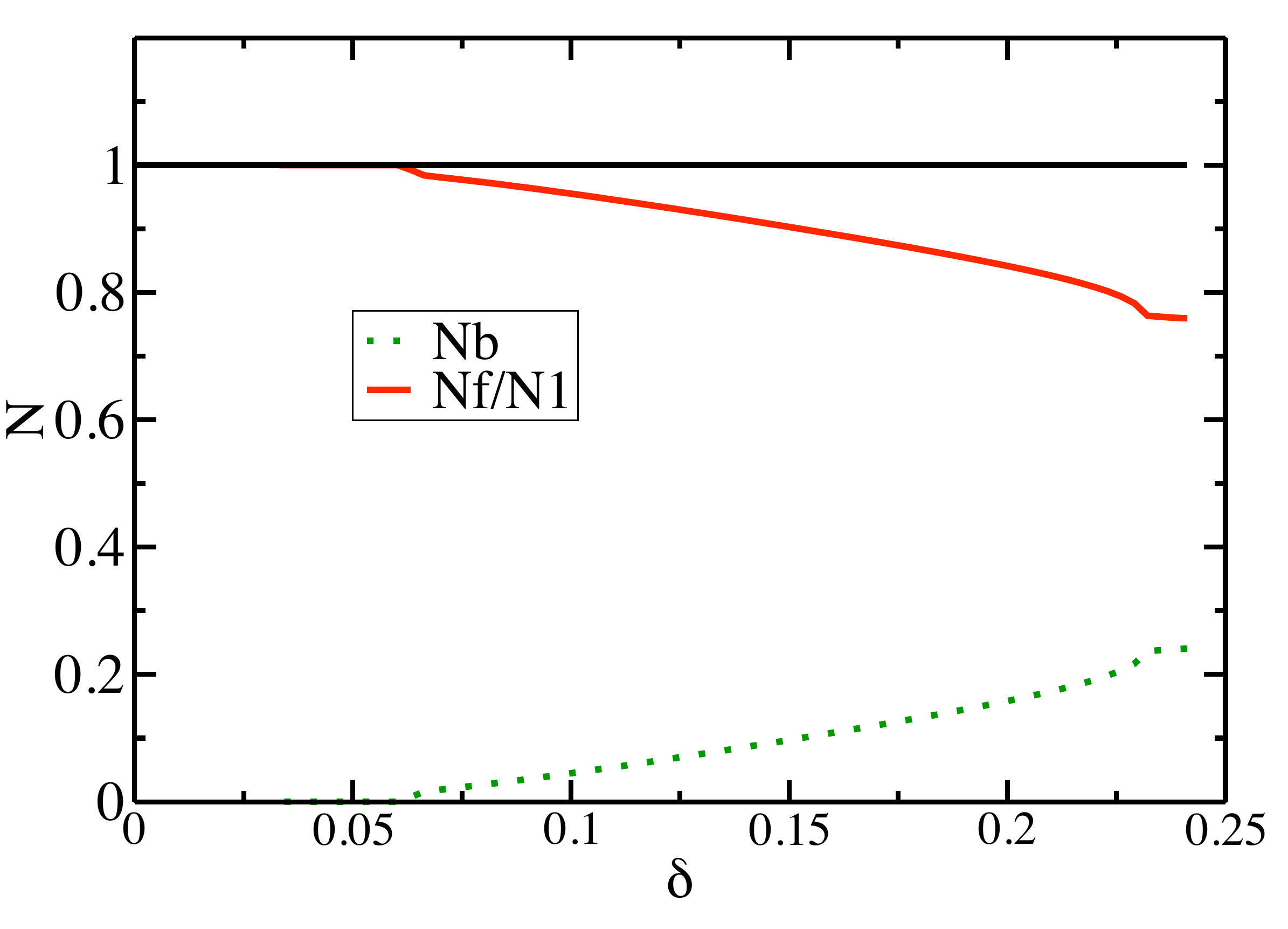}
\caption{ The number of bosons $n_{b}$ (green online) and of the f-fermions (red online) per site in the system as well as their sum (black). The local constraint  of no double occupancy is preserved throughout the explored coverage range. } \label{constraint}
\end{figure}

In the  Figure \ref{constraint} we have plotted directly the  number of holons in the hybridized phase, as given from our mean-field theory.  The number of holons determines the number of  holes in the first layer as compared to the value at half-filling.
We can see that although the order of magnitude is correct close to the QCP, far away from it we obtain some values of $n_b$ too big from what is observed experimentally. In particular,  it is believed that close to the coverage corresponding to the promotion of the second layer,  the number of holons should decrease  so that the number of f-fermions in the first layer should be again close to half-filling. We don't observe any hint of this decreasing. It shows that the domain of validity of our model is close to the QCP. Far away from it, we miss the physics of  exhaustion \cite{nozieres,burdin2}, where there are not enough free fermions in the second layer to Kondo screen the many f-fermions in the first layer.

    \label{sec:MF}
    
    \section{Fluctuations}
    \label{sec:fluc}
\begin{figure}[h] 
    \centering
    \includegraphics[width=8 cm]{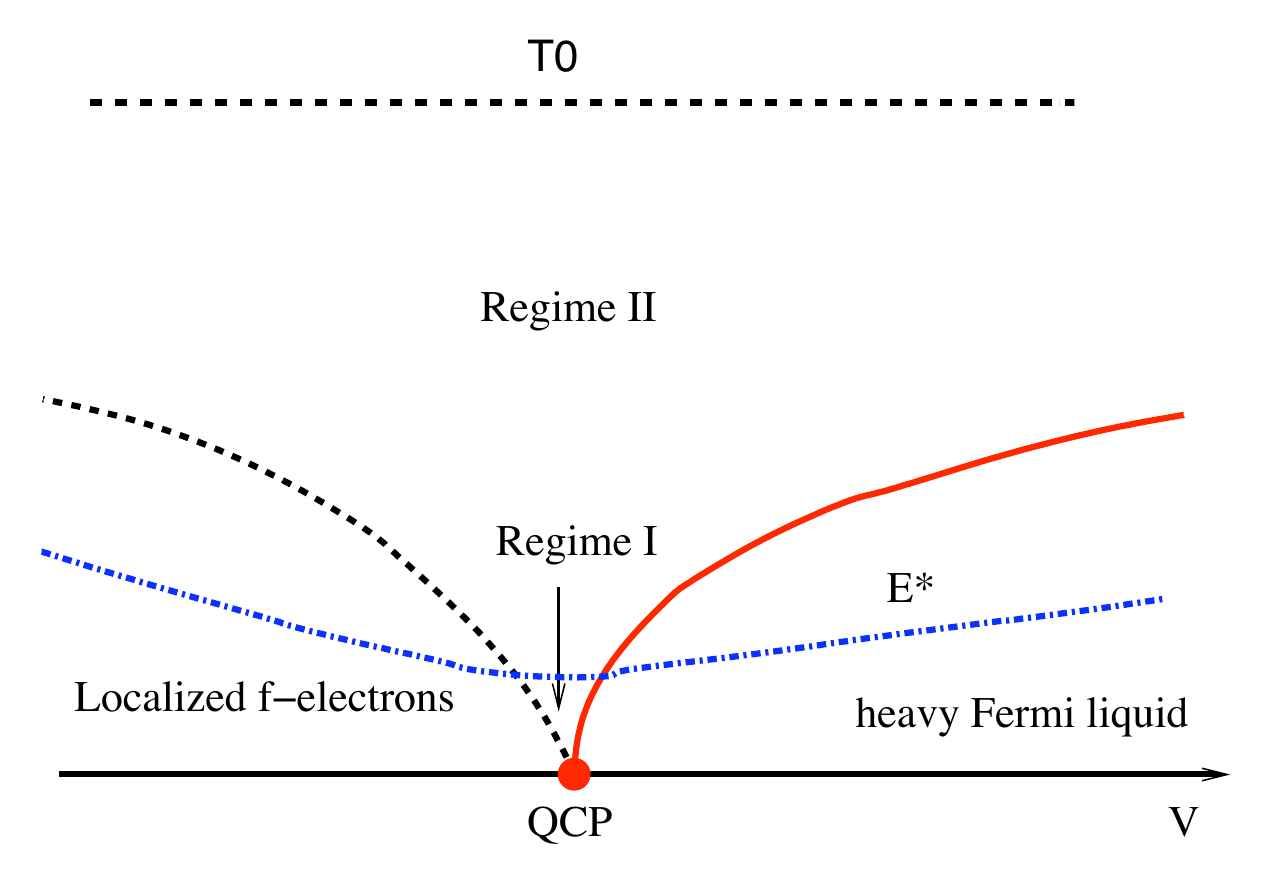} 
    \caption{  Schematic for the Kondo breakdown QCP in the Anderson lattice\cite{cathlong}. On the left, where the holons are not condensed is the localized phase. On the right is the heavy Fermi phase. The QCP is multi-scale ; for $T \leq E^*$, the dynamical exponent is $z=2$ and for $T \geq E^* $ it is $z=3$. }
    \label{phase-diag} 
    \end{figure}

   In what follows we will be interested in fitting the experimental data.  We identify the regime of critical fluctuations  experimentally accessible with the higher energy regime of the order parameter (see Fig. \ref{MF}).  Within our theory, we are situated in the intermediate regime around the Kondo breakdown  QCP, i.e. the regime for which  the dynamical exponent $z=3$. We refer the  reader to previous studies of  the Kondo breakdown for more details \cite{us,cath,cathlong,uslong}.   To give a small summary of the situation (see Fig. \ref{phase-diag}), the main finding of the Kondo breakdown QCP is its multi-scale character. There exists an energy scale $E^*$  differentiating two regimes. In the low temperature regime we have the dynamical exponent $z=2$ \cite{dyn}, with no damping. In the high temperature regime, we have the exponent $z=3$ an the bosonic mode corresponding to the fluctuations of the order parameter is over-damped by  the particle-hole continuum.   In this paper we focus on the $z=3$ regime, arguing that  $E^*$ is very small in this system.
    Indeed, from the theory (see for example \cite{uslong}) we know that $E^* \simeq 0.1 (q^*/q)^3 T_K$, with $q^*$ the mis-match of the two Fermi surfaces at the QCP. Here  $T_K$ can be taken as  the typical energy scale of the system which is  typically of the order of  $ T_K = 100 \ mK$.  At the QCP, we evaluate $q^*/k_F$ which is 
    \bea
    q^*/k_F &  = &  1- k_c/k_F \ , \nonumber \\
                & = & 1- ( 6.3 / 9.95)^{1/2}  \ , \nonumber \\
                 & = & 0.2 \ . \nonumber \eea Hence  we obtain
                 \bea
                 E^* & =  & 8. 10^{-4} T_K  \ , \nonumber \\
                  & = & 8. 10^{-5} K \ , \eea which is a too small energy scale to be accessible experimentally for this set-up.

The holon propagator in the intermediate regime (z=3) reads:
 \beq \label{eqn3}
 D^{-1}_b ( q, \Omega_n ) = D_0^{-1} \left [  q^2 + \xi^{-2}  +  \frac{\gamma |\Omega_n | }{ \alpha^\prime  q } \right ] \  ,  \eeq

with $D_0 = 4 k_F^2 / (\rho_0  V^2 ) $, 
  $ \gamma =  m V^2 D_0 / ( \pi v_F ) $  $\alpha^\prime = b^2 \alpha + J/t$, $ \rho_0 = m_c/  ( 2 \pi )$ is the
  c-fermions density of states and $\xi$ is the correlation length, associated with the fluctuations of b, given by $\xi^2 = D_{0}^{-1}m_{b}^{-1}$ where $m_{b}$ is the holon mass at $T=0$.

\subsection{The Holon mass}

The static part of the holon mass is evaluated by differentiating twice the mean-field energy (\ref{MFFE}) with respect to the holon field $b$ given the contraints (\ref{MFE}). One finds

\bea
m_{b} & = &  2 b T \sum _{\mathbf{k}, \omega} \left [ \alpha \epsilon_{\mathbf{k}} \frac{\partial G_{ff}}{\partial b} + V^2 \frac{\partial P_{fc}}{\partial b} \right ] \,\,. 
\eea
The summation over $(\mathbf{k}, \omega)$ is evaluated anatically for a linearized dispersion bandwidth at $T = 0$ and the result is given in Appendix A.
\newline
\newline

The temperature dependence of the holon mass is computed by evaluating the corrections to scaling to the boson propagator.  There are two types of corrections to scaling. One  contribution is the renormalization of the boson propagator  coming  to their coupling to the fermion loops.
 \[  \parbox[c]{9cm}{\includegraphics[width=9cm]{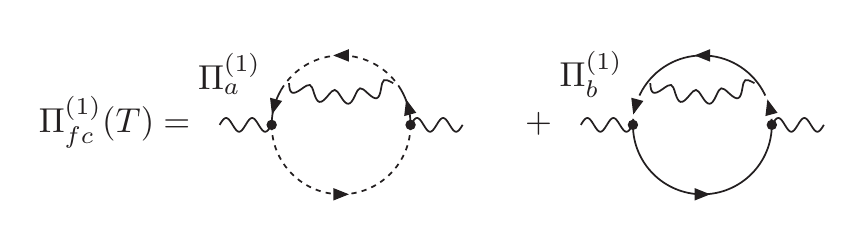}} \]
 This type contribution was first evaluated close to a QCP in Ref.\cite{belitz}.
   We first note that the two diagrams are proportional: $ \Pi_a^{(1)} = \alpha^\prime \ \Pi_b^{(1)}$ and that there is no corresponding vertex insertion at the first order.  Hence, although the QCP occurs in the charge channel, we have no cancellation of this set of diagrams. This is in deep contrast to what occurs close to a ferromagnetic QCP or in the theory of non analytic corrections to the Landau Fermi liquid, where this set of diagrams  cancels in the charge channel \cite{maslov-chub}. This type of diagram is known to be dangerous, and carries a minus sign, which destabilizes the fixed point.  The diagram for $\Pi_{a}^{(1)}$ is computed in the intermediate energy regime with the dynamical exponent $z=3$.

 On the other hand, we have the direct  mass renormalization coming from the standard $\phi^4$-type corrections to scaling, which has to opposite effect of stabilizing the fixed point
  \[  \parbox[c]{2cm}{\includegraphics[width=2cm]{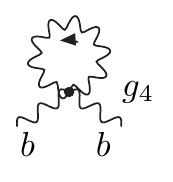}}. \] Here $g_4$ comes from the quartic term of the holon action derived from (\ref{MFFE}) in a Ginzburg-Landau approach and contains the ferromagnetic short range correlations $J$ ; we find $g_4 = -J/4 + V^4/(2 \alpha^{'2}D^3)$.  We have as well as the correction  to the boson mass coming from the gauge fluctuation, which stabilizes as well the QCP, but is subdominant compared to the two previous ones.
   \[  \parbox[c]{3cm}{\includegraphics[width=3cm]{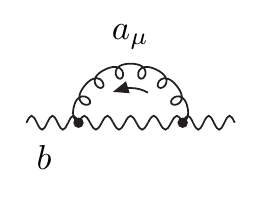}} \ .  \]
   
Summing the dominant diagrams (see Appendix \ref{pia1}) yields a logarithmic correction to scaling

\beq \label{eqn5}
 m_b(T) = m_b (T=0)  + C   \  T Log T  \ , \eeq

where $C$ had to be adjusted to $C = 7.5 \ 10^{-3}$ to fit the data, while the analytic evaluation gives  $$C= \left [  \frac{(1+\alpha^\prime)}{8} - \frac{1}{3} \right ] \frac{ V^2}{8 (\alpha^\prime D)^2} + \frac{  D J}{6 V^2}$$
\newline
The balance of the two contributions in favor of the $g_4$ coupling ensures the stabilty of the fixed point.   

\subsection{The effective mass}

The effective mass $m^*$ is determined from the free energy of the system by :
$$F = - \frac{\pi T^2}{6}m^*$$
The free energy is evaluated using the Luttinger-Ward functional \cite{maslov-chub}
\beq F = F_{MF} +  T/2 \  \sum_n  \int d^2 q/ (2 \pi)^2  \log \left [ D^{-1} (q, \Omega_n ) \right ], \label{LW}\eeq
where $F_{MF}$ is the free energy at the mean-field (\ref{MFFE}) and $D(q, \Omega_n)$ is the full propagator of the holons.
Note that we have neglected the role of the gauge fields in this formulation, because the re-normalization of the effective mass is to be evaluated inside the ordered phase where the gauge fields are gapped through the Higgs mechanism. At the mean-field level, the system consists of the upper and lower bands, we get then : 

\beq \label{eqn4}  
 F = - \frac {\pi T^2}{6} \left [ 2 \pi ( \rho_+ + \rho_- ) + \frac{ \gamma \xi}{ 4 \alpha^\prime} \right ] \ , \eeq
where $\rho_{+}(\rho_{-})$ is the density of states at the Fermi surface of the upper (lower) band given by :
$$\rho_{\pm}= \rho_{0}\left (\frac{\partial E_{\mathbf{k}\pm}}{\partial \epsilon_{\mathbf{k}}}\right )^{-1}_{|E_{\pm}=0}$$
The calculation is done in Appendix \ref{app:FE}. 
\newline

The effective mass reads directly
$$m^* = 2 \pi ( \rho_+ + \rho_- ) + \frac{ \gamma \xi}{ 4 \alpha^\prime}$$
\begin{figure}[ht]
\includegraphics[width=8.5cm]{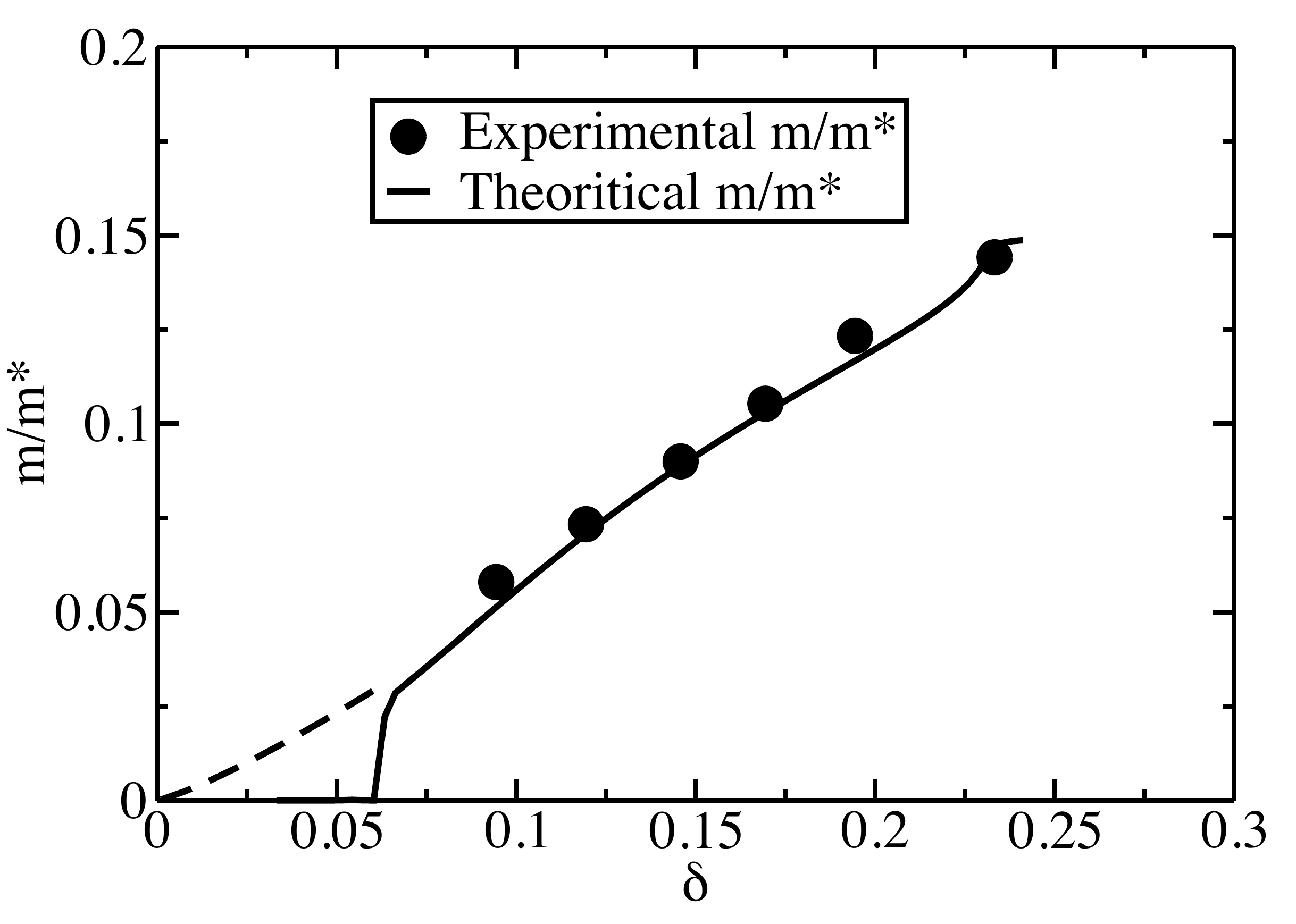}
\caption{The inverse effective mass $ m/ m^*$
in the Anderson lattice model for the
He$^3$ bi-layers\cite{footnote}. The dots are experimental data from\cite{saunders}. 
The fitting parameters for this model are detailed in the text \cite{adel}.} \label{mms}
\end{figure}

The result for $m^*$ is shown in Fig. \ref{mms} where it is compared to the results of experiment \cite{saunders}. We see that the inverse effective mass follows the same behaviour as the order parameter and vanishes at the theoritical QCP. Here again, if we extrapolate the high energy regime down to zero temperature, we can identify a fictious point where the effective mass could vanish, if it has not its peculiar behaviour into two regimes. This extrapolation is linear and follows closely the one found by the experimentalists.
\subsection{The coherence temperature}
\begin{figure}[h]
\includegraphics[width=8.5cm]{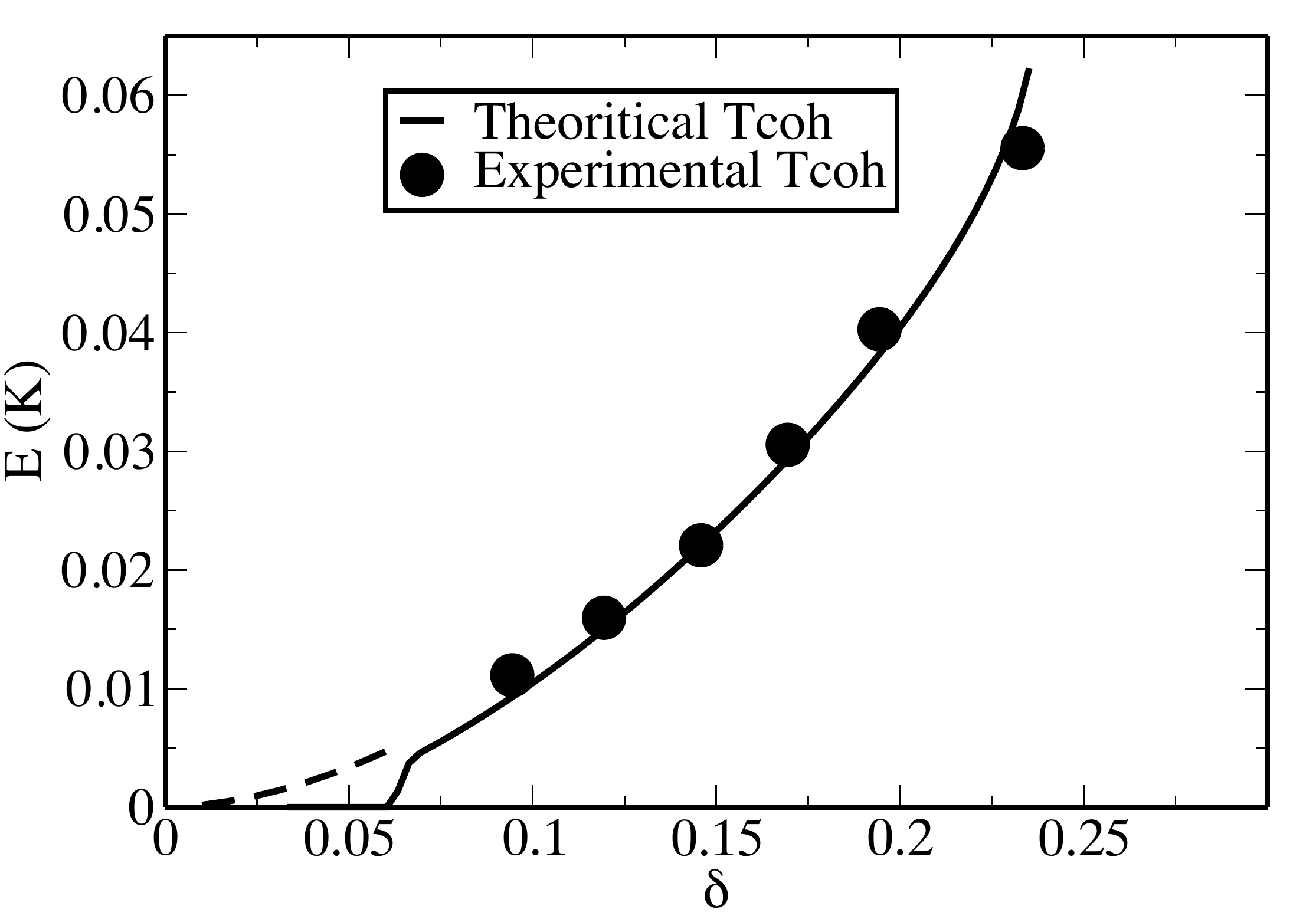}
\caption{The coherence temperature $T_{coh}$
in the Anderson lattice model for the
He$^3$ bi-layers. The dots are experimental data from\cite{saunders}. 
The fitting parameters for this model are detailed in the text \cite{adel}.} \label{Tcoh}
\end{figure}

The coherence temperature is defined by the cross-over condition 
$$m_{b}(T_{coh})=0,$$
where $m_{b}(T)$ is the temperature dependant holon mass, given in (\ref{eqn5}).
\newline
\newline
The equation is solved numerically, using the results found for the order parameter b in the previous section , and the result is plotted in Fig. \ref{Tcoh}.
\newline

The coherence temperature has the same qualitative behavior : it vanishes at the real QCP and we can extrapolate its high energy regime down to zero temperature closely to a quadratic power law in unit coverage $\delta$.
\newline 

In fact, the exponents of the effective mass and the coherence temperature can be understood in a simple way.
 For $z=3$ theories in the Fermi liquid phase, the effective mass goes like the correlation length \cite{jerome}  $ m/ m^* \sim \xi^{-1}$. From the 
 dispersion of  the
  boson mode we see that $\xi^{-1} \sim \sqrt{m_b} \sim b $.  Now  the coherence temperature goes like $b^2$. In the regime where $b$ varies linearly with 
  the coverage
   $n$ we thus get
  \beq \label{eqn6}
  \begin{array}{ll}
  m/m^* \sim cst - n \ , \; \; \; \; & \; \; \; T_{coh} \sim  (cst - n )^2 \  . \end{array} \eeq   

    \section{Discussion}
    \label{sec:discussion}
    
    One of the main  general observation one gets from the experimental data is the asymmetry of the phase diagram, as far as the quantum fluctuations are concerned. Indeed the increase of the effective mass appears only  from the right of the phase diagram which corresponds to low doping (see Fig. \ref{mms}). From the left of the phase diagram the fluctuations seem to be frozen out. 
    \newline 
    
    Another observation is the  quasi-absence of quantum critical (QC) regime in temperature for this system, unlike for the heavy fermions. Indeed a Curie law for the spin susceptibility is observed at very low temperatures in the localized phase and directly above $T_{coh}$ in the hybridized phase, indicating that the system very quickly goes into a regime  of free spins, hence missing the usual quantum critical regime typical of QCP.
    \newline
    
    The key to understanding these two observations is that in this system the energy scales are completely different form the ones that appear in heavy fermion systems. 
 \begin{figure}[ht]
\includegraphics[width=7.5cm]{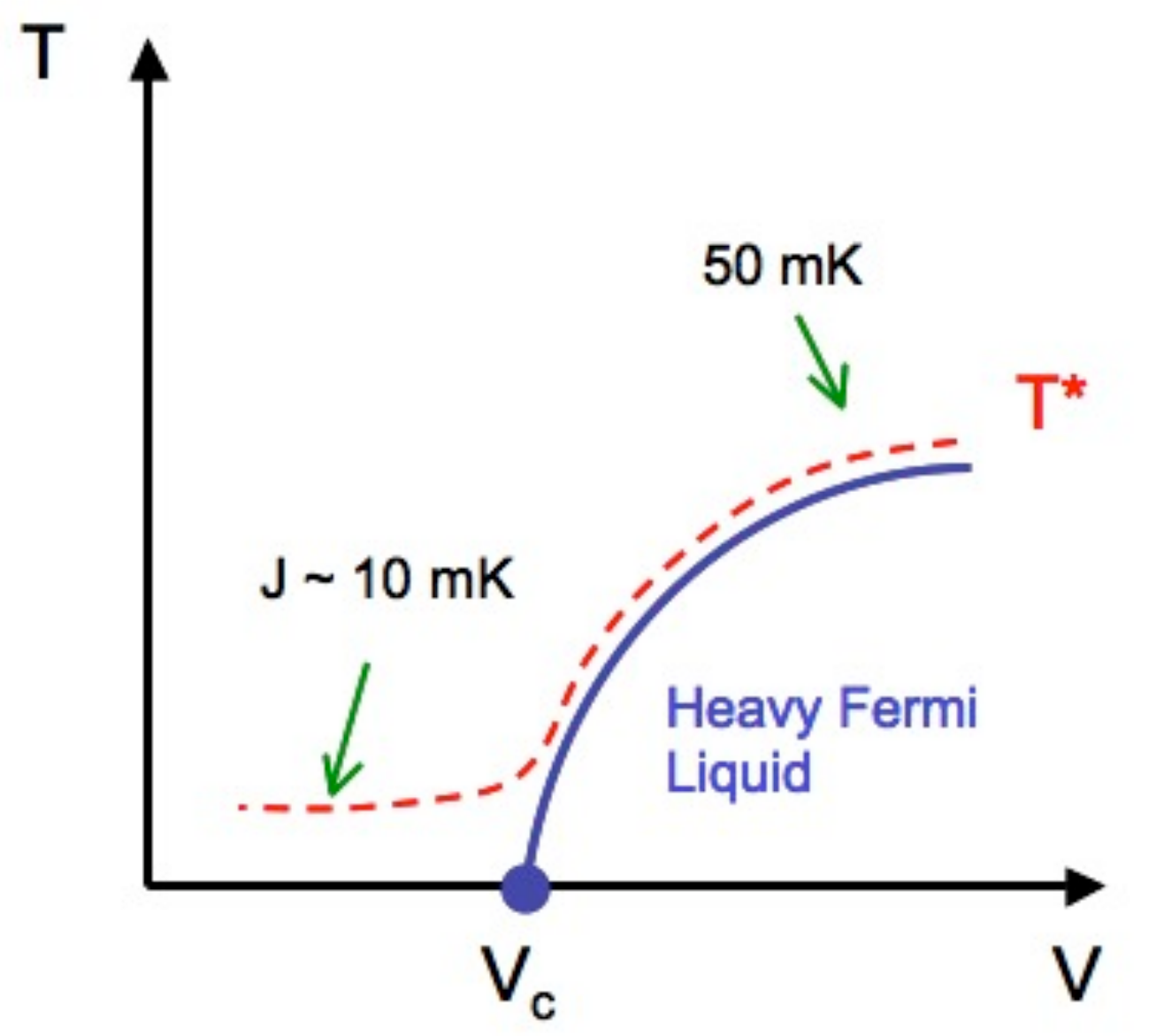}
\caption{Asymmetry of the phase diagram. Above $T^*$ the entropy $R Ln 2$ is released. There are two ways of quenching the entropy, first through the formation of the heavy Fermi liquid phase where the hybridization is non zero ( on the right of the phase diagram), second through the formation of the spin liquid (on the left of  the phase diagram).} 
\label{asymmetry}
\end{figure}
\newline

The Curie law is observed when the entropy $R \ln{2}$ is released, above a characteristic temperature $T^*$. In our model, two mechanisms are responsible for quenching the entropy, namely the formation of the spin liquid and of the  heavy Fermi liquid.   $T^*$ is thus determined by the relative strength of these two mechanisms. Technically, $T^*$ is by the first irrelevant operator of the theory. We see in Figure \ref{asymmetry} that on the  left side of  the phase diagram, the main quenching mechanism corresponds to the formation of the spin liquid, while on the right side of the phase diagram, the two mechanisms coincide and are roughly of the same strength. The asymmetry of the phase diagram can thus be accounted for, in this model, by the fact that on the localized side (left side)  the spinons' bandwidth, which determines the scale  of the the formation of the spin liquid, is typically given by the value of the exchange parameter $ J \simeq 7 \ mK $. Alternatively, in the  hybridized phase, the bandwidth  of the spinons is enlarged, due to the holon fluctuations $  D_f = J + n_b  \alpha D $. This increase of the bandwidth in the hybridized phase is typical of a slave-boson description of a Mott transition \cite{lee-review}.

\begin{figure}[ht]
\includegraphics[width=7.5cm]{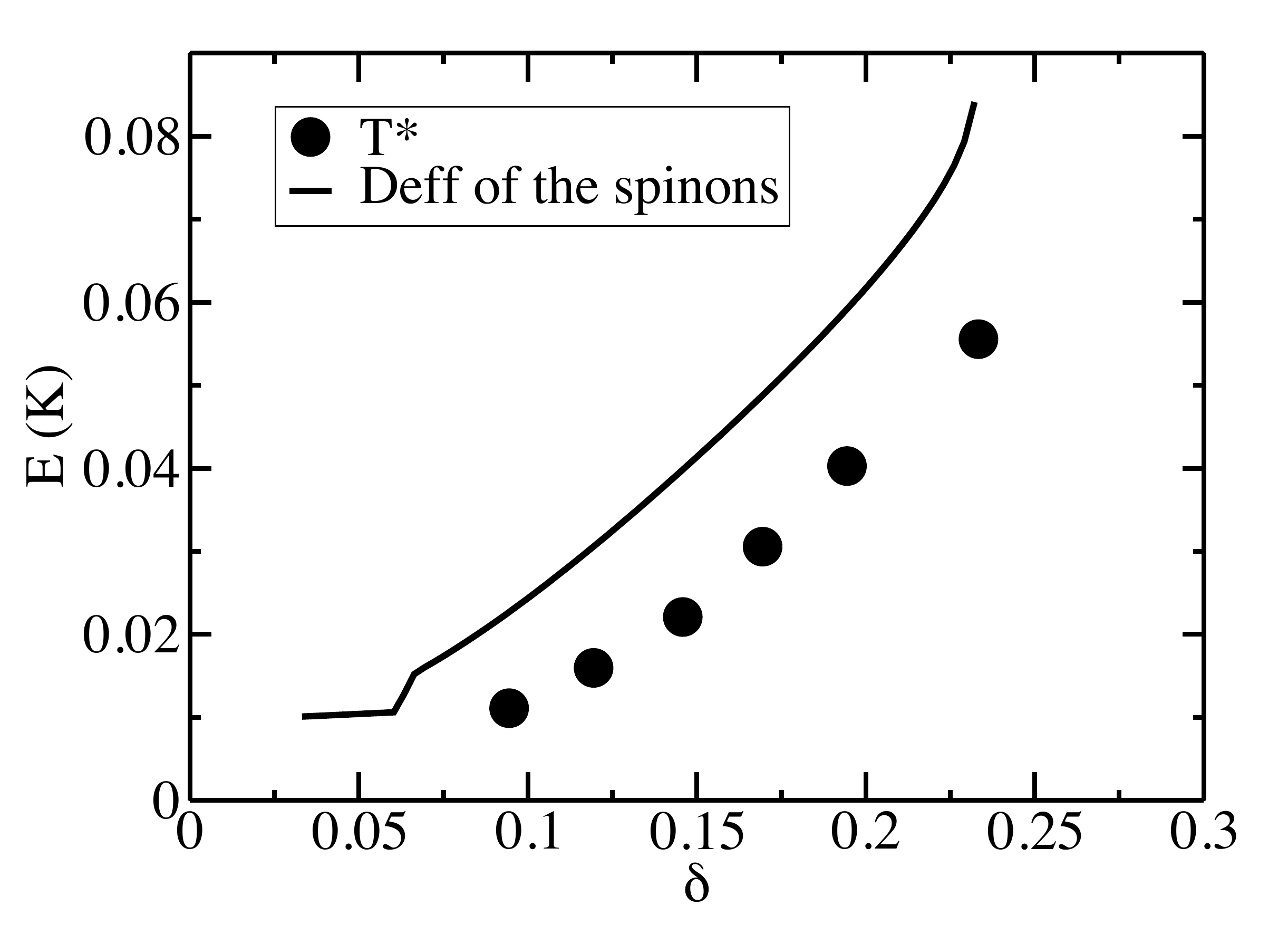}
\caption{ The effective bandwidth of the spinons $D_f = J+\alpha D n_b$ and the experimental characteristic temperature $T_0.$} \label{spinon_bandwidth}
\end{figure}

 In the hybridized phase the coincidence, within the experimental uncertainties (between 5 and 10 mK\cite{saunders}), in energy between the cross-over coherence temperature and the effective mass of the spinons (See Fig. \ref{spinon_bandwidth}) explains that the quantum critical regime is quenched, the free spin behavior being admittedly quickly observed above the temperature which delimits the upper-critical regime. 
\newline

\section{Conclusions}
\label{sec:conclusions}
In the present article, we give the details of calculation whose results have been presented in a previous Letter \cite{adel}. The system studied, He$^3$ bilayers, is one of the simplest physical ones, with negligible spin-orbit interaction and no crystal-field interactions, to show quantum criticality (QC) similar to the one observed in complicated intermetallic heavy fermions compounds. 
\newline

Using the Kondo-Breakdown\cite{us,cath,cathlong} scenario of an itinerant QCP, we examine the possible origine of the QC observed experimentally as fluctuations of an effective hybridization. The theoritical model is an extended version of the Anderson lattice model with a dispersion of the f-fermions and inter- and intra- Coulomb repulsion.
\newline

We benefited from the extensive literature on He$^3$ to extract carefully most of the parameters of the model from  the bare parameters. Crucial parameters, like the hybridization, were used as fitting parameters owing to the level of approximation of our study. Finally, we have emphasized some differences with intermetallic heavy fermions compounds.
\newline

We were successful enough to account for most of the experimental features. First,  we have explained why there are seemingly two apparent QCPs which fit at the right respective coverage. The experimental one results from an extrapolation to zero temperature of an intermediate energy regime, while the theoritical one charachterizes the vanishing of the effective hybridization. We reproduced then the  slopes  {\it and} exponents of the  coherence temperature and effective mass closely to the experimental results. The apparent lack of quantum critical behavior in temperature is qualitatively explained by the remarkably low energy scale of the spin liquid parameter on the ordered side, and the coincidence between the coherence temperature and the effective mass of the spinons in the hybridized one. Finally, we recover the fact that the activation gap, observed experimentally, has to vanish in the Fermi liquid phase  before the critical coverage is reached, right when the system enters a strong hybridization regime for which the upper hybridized band becomes empty.
\newline
  
Our study suffers though from some weakness and drawbacks. We used 4 fitting parameters, 3 for the hybridization and one for the slope of the coherence temperature. This is expected in any mean-field appraoch, in particular owing to the crucial role of the hybridization for the Kondo breakdown QCP, and can not be avoided at this level. The fact that the number of holons $n_b$ is too big away from the QCP, especially near the promotion coverage of the second layer, restricts the domain of validity of our model very close to the QCP. Finally,  magnetism on the ordered side of the phase diagram is not handled in our model. Magnetism is best considered in the so-called slave fermions approach, which in turn describes badly the hybridized phase.  

But still the model is simple and strong enough  to make predictions and put them to the test. This is the only proposed model of a de-confined QCP to be tested from ab-initio parameters. 
\newline

 
 Useful discussions with H. Godfrin, G. Misguich, M. Neumann, J. Nyéki, O. Parcollet, M. Ferrero and J. Saunders are aknowledged.
\newline

 This work is supported by the French National Grant ANR26ECCEZZZ.
\newpage

\appendix
\section{ Evaluation of some integrals}
\label{app:MF}

   In here, we will evaluate the integrals in the mean-field equations (\ref{MFE}). At $T=0$, the calculation of these integrals is analytical for linearized bands in which case :
\beq\sum _{k} \rightarrow \rho_{0} \int^{D}_{-D} d\epsilon ,\label{lin}\eeq
where $\rho_{0}$ is the density of states at the fermi surface.
\newline

Let's call 
  \bea  \label{appendeq1}
  {\cal A}  & = &   T \sum _{\mathbf{k}, \sigma,  \omega_n }  G_{ff} (\mathbf{k} , i \omega_n)  \nonumber ,  \\
  {\cal B } & = &  T \sum _{\mathbf{k} \sigma, \omega_n } P_{fc}   ( \mathbf{k} , i \omega_n )     \nonumber , \\
  {\cal C} & = &  T \sum_{\mathbf{k}, \sigma, \omega_n } \epsilon_\mathbf{k}  G_{ff}( \mathbf{k}, i \omega_n ) \  ,   \nonumber \\
{\cal D} & = & T \sum_{\mathbf{k}, \sigma, \omega} \alpha \epsilon_{\mathbf{k}} \frac{\partial G_{ff}}{\partial b} , \nonumber \\
{\cal E} & = & T \sum_{\mathbf{k}, \sigma, \omega} \frac{\partial P_{fc}}{\partial b} \nonumber
 \eea
   We diagonalize the $2x2$ matrix which accounts for the hybridization of the f- and c- bands:
   \bea 
   E_{\mathbf{k}\pm} & = &  \frac{1}{2} \left [ \epsilon^0_\mathbf{k} + \epsilon_\mathbf{k} \pm \sqrt{\Delta} \right ]\nonumber , \\
   \Delta & = & ( \epsilon^0_\mathbf{k}  - \epsilon_\mathbf{k} ) ^2 + 4 ( b V ) ^2  \  . \nonumber \eea
   The integrals are all performed in the same way, first by summing over the Matsubara frequencies , and second by  doing the momentum integration.  The momentum integration is done  by linearization of  the band.
   \bea \label{appendeq2}
   {\cal A} & = &  2 T \sum _{k, \omega_n} \frac{ ( i \omega_n - \epsilon_\mathbf{k} ) }{  ( i \omega_n -E_{\mathbf{k}-} ) ( i \omega_n - E_{\mathbf{k}+} ) }  \nonumber \\
     & = & 2 \rho_0 \int_{-D}^{D} d \epsilon  \int \frac{ - n_F (z) } { 2 i  \pi } \ \frac{ (z - \epsilon) }{(z - E_{-}) (z-E_{+})}  \ d z \ , \nonumber \\ 
     & &\mbox{ where the contour is on the whole complex plane } \nonumber \\
     & = &   2 \rho_0 \int_{-D}^{D} d \epsilon \left ( \frac{n_F ( E_{-})  ( E_{-} - \epsilon _k ) }{ ( E_{-} - E_{+} ) }  - \frac{n_F(E_{+}) ( E_{+} - \epsilon_k) }{( E_{-}- E_{+} )} \right )  \nonumber \\
     & = &   \rho_0 \int_{-D}^{\epsilon_m} d \epsilon \frac{ - y + \sqrt{ y^2 + 4 ( b V  )^2 }} {  \sqrt{ y^2 + 4 ( b V  )^2 }}  \nonumber \\
     & - &  \rho_0 \int_{-D}^{\epsilon_p} d \epsilon \frac{ - y - \sqrt{ y^2 + 4 ( b V  )^2 }} {  \sqrt{ y^2 + 4 ( b V  )^2 }} \nonumber \  , 
     \eea  with  $ \epsilon_m $ and $\epsilon_p $ the    Fermi levels for the upper and lower bands respectively.
     \bea \label{appendeq3}
     \epsilon_m & = &   (- \epsilon _f + \alpha^\prime \mu  - \sqrt{ ( \epsilon_f + \alpha^\prime \mu )^2 + 4 \alpha ^\prime  ( b V  ) ^2 }  )   / ( 2 \alpha^\prime )  \nonumber \\
       \epsilon_p & = &   ( - \epsilon _f + \alpha^\prime \mu  + \sqrt{ ( \epsilon_f + \alpha^\prime \mu )^2 + 4 \alpha ^\prime  ( b V ) ^2 }  ) /  ( 2 \alpha^\prime  )   \nonumber \\
       && \mbox{ with the conditions } \begin{array}{ll} -D \leq \epsilon_m \leq  0 ; \; \; & 0 \leq \epsilon_p \leq D  \ , \end{array}  \nonumber \\
       && \mbox{and} \;  \alpha^\prime =  \alpha b^2 + \phi_0/ D \ . \nonumber \eea
       One obtains
       \bea {\cal A } &  =  &  \frac{  \rho_0}{  ( 1 - \alpha^\prime) } \left ( - 2 y_{-D} + y_m - \sqrt{ y_m^2 + 4 ( b V )^2} \right . \nonumber \\
       &  + & \left .  y _p + \sqrt{ y_p^2 + 4 ( b V  )^2}  \right )  \  ,  \nonumber \eea
      \[ \begin{array}{ll} \mbox{ where } \; & \begin{array}{l} 
       y_m   =   ( 1 - \alpha^\prime  ) \epsilon_m - \epsilon_f - \mu ;  \\
        y_p   =   ( 1 - \alpha^\prime  ) \epsilon_p - \epsilon_f - \mu ;  \\
         y_{-D}   =   - ( 1 - \alpha^\prime  ) D- \epsilon_f - \mu \  . \end{array} \end{array} \]
         
 We proceed in the same way for ${\cal B } $, ${\cal C} $, ${\cal D} $ and ${\cal E} $  to find:
 \[
 {\cal B } = \frac{ 2 \rho_0} { ( 1 - \alpha^\prime) }  \ln{ \left [ \frac{ y_m + \sqrt{ y_m^2 + 4 ( b V  )^2}}{y_p + \sqrt{ y_p^2 + 4 ( b V  )^2} }\right ]} \ . \]
 
 \bea 
{\cal C }   & = &  \frac{  \rho _0}{  ( 1 - \alpha^\prime ) ^2 } \left [  - 2 ( \epsilon_f + \mu ) ( y _{-D} )  + y_{-D}^2 \right . \nonumber \\
  & +  & 2  (b V )^2 \ln{ \left (  \frac{ y_m + \sqrt{ y_m^2 + 4 ( b V  )^2}}{y_p + \sqrt{ y_p^2 + 4 ( b V  )^2} }\right ) }  \nonumber \\
 & + &  ( \epsilon_f + \mu ) y_m + y_m^2 \nonumber \\
 & - & ( y_m/2 + \epsilon_f + \mu ) \sqrt{ y _m ^2 + 4 ( b V  )^2 }  \nonumber \\
  & + &   ( \epsilon_f + \mu ) y_p + y_p^2 \nonumber \\
  &  + & \left .  ( y_p/2 + \epsilon_f + \mu ) \sqrt{ y _p ^2 + 4 ( b V  )^2 } \right ] \  . \nonumber \eea

\bea
{\cal D} & = - & \frac{4 \alpha   V^2 b \rho _0}{  ( 1 - \alpha^\prime ) ^2 }\ln{ \left [ \frac{ y_m + \sqrt{ y_m^2 + 4 ( b V  )^2}}{y_p + \sqrt{ y_p^2 + 4 ( b V  )^2} }\right ]} \nonumber
\eea

\bea
{\cal E} & = & \frac{4V^2 b \rho_{0}}{(1-\alpha^\prime)} \left [ \frac{y_{p}}{2 V^2 b^2 \sqrt{y_{b}^{2}+ 4 V^2 b^2}} \right . \nonumber \\
& - & \left . \frac{y_{m}}{2 V^2 b^2 \sqrt{y_{m}^{2}+ 4 V^2 b^2}} \right . \nonumber \\
& + & \left . \frac{-2}{(y_{m}-\sqrt{y_{m}^{2}+ 4 V^2 b^2})\sqrt{y_{m}^{2}+ 4 V^2 b^2}} \right . \nonumber \\
&+& \left .  \frac{-2}{(-y_{p}-\sqrt{y_{p}^{2}+ 4 V^2 b^2})\sqrt{y_{p}^{2}+ 4 V^2 b^2}} \right ] \nonumber
\eea

\section{Corrections to scaling for the holon mass}
\label{pia1}
In this appendix, we discuss the stability of the QCP. We will start by evaluating the diagram $\Pi_{a}^{(1)}$
\[  \parbox[c]{8cm}{\includegraphics[width=7cm]{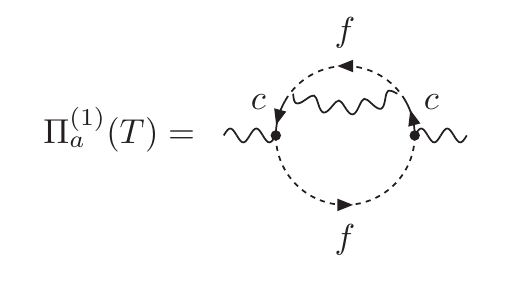}} \]

\bea
 &  \Pi_a^{(1)} (T)&  =   2  T^2 V^4 \sum_{n,m \neq 0} \sum_{\mathbf{k}, \mathbf{q}}   D_b(\mathbf{q}, \Omega_m ) \nonumber \\
 & &\times \ G_c^2(\mathbf{k},\omega_n)  G_f (\mathbf{k}, \omega_n) G_f(\mathbf{k+q}, \omega_n + \Omega_m)  \ .  \nonumber  
\eea
Introducing the angle $\theta$ defined by $\epsilon_{\mathbf{k+q}}=\epsilon_{\mathbf{k}} + v_{F}q\cos{\theta}$ and considering linearized bands like in (\ref{lin}), we have, with $\bar g = 8k_{F}^2 V^2/\rho_{0}$
\bea
& \Pi_a^{(1)} (T) & =   \bar g \rho_{0} T^2 \sum_{n,m \neq 0} \sum_{\mathbf{q}} \int d \,\, \theta d \epsilon  \frac{1}{q^2   +  \frac{\gamma |\Omega_m | }{ \alpha^\prime  q }} \nonumber \\
 & &\times \frac{1}{ ( i \omega_n  - \epsilon + \mu )^2} \ 
   \frac{1}{ ( i \omega_n  - \alpha^\prime \epsilon - \epsilon_f ) }   \nonumber  \\
 & & \times \frac{1}{ ( i \omega_n  + i \Omega_m   - \alpha^\prime \epsilon - \alpha^\prime v_F q \cos{\theta} - \epsilon_f ) } \nonumber 
   \eea 

Summing over the fermionic Matsubara frequencies $\omega_{n}$ then integrating over $\epsilon$, we get
\bea
&\Pi_{a}^{(1)}& = \bar g \rho_{0} \alpha^\prime T \sum_{\mathbf{q}, m \neq 0} \int d \theta \frac{i \Omega_{m}}{q^2   +  \frac{\gamma |\Omega_m | }{ \alpha^\prime  q }} \nonumber \\
&& \times \frac{1}{i \Omega_{m} - \alpha^\prime v_{F}q \cos{\theta}} \nonumber \\
&& \times \frac{1}{i \alpha^\prime \Omega_{m}- \alpha^\prime v_{F}q \cos{\theta}-\alpha^\prime \mu - \epsilon_{f}}\nonumber \\
&& \times \frac{1}{i \Omega_{m}- \alpha^\prime v_{F}q \cos{\theta}-\alpha^\prime \mu - \epsilon_{f}}  \nonumber 
\eea
Now, we have $\mathbf{q}=(q_{x}, q_{y})$ with $q_{x}=\cos{\theta}$ and $q_{y}=q \sin{\theta}$. We suppose  $q_y \gg q_x$ and expand $ q= \sqrt{q_x^2+q_y^2} \simeq | q_y| + q_x^2 / (2 | q_y| ) $. We find
       \bea
 &  \Pi_a^{(1)} (T)&  = \frac{\bar g \rho_{0} \alpha^\prime}{4\pi^2} T \sum_{m \neq 0}\int dq_x \int_{|q_x|}^{\Lambda}  \ dq_y  \frac{i \alpha^{\prime} \Omega_m}{\gamma |\Omega_m | }  \nonumber \\
 & &\times \frac{ \left (  | q_y| + q_x^2 / (2 | q_y| ) \right )}{i \Omega_{m} - \alpha^\prime v_{F}q_{x}} \ ; \nonumber\\
&& \times \frac{1}{i \alpha^\prime \Omega_{m}- \alpha^\prime v_{F}q_x-\alpha^\prime \mu - \epsilon_{f}}\nonumber \\
&& \times \frac{1}{i \Omega_{m}- \alpha^\prime v_{F}q_x-\alpha^\prime \mu - \epsilon_{f}} , \nonumber  \eea
where $\Lambda$ is an ultra-violet cut-off.

\begin{figure}[h]{\includegraphics[width=8cm]{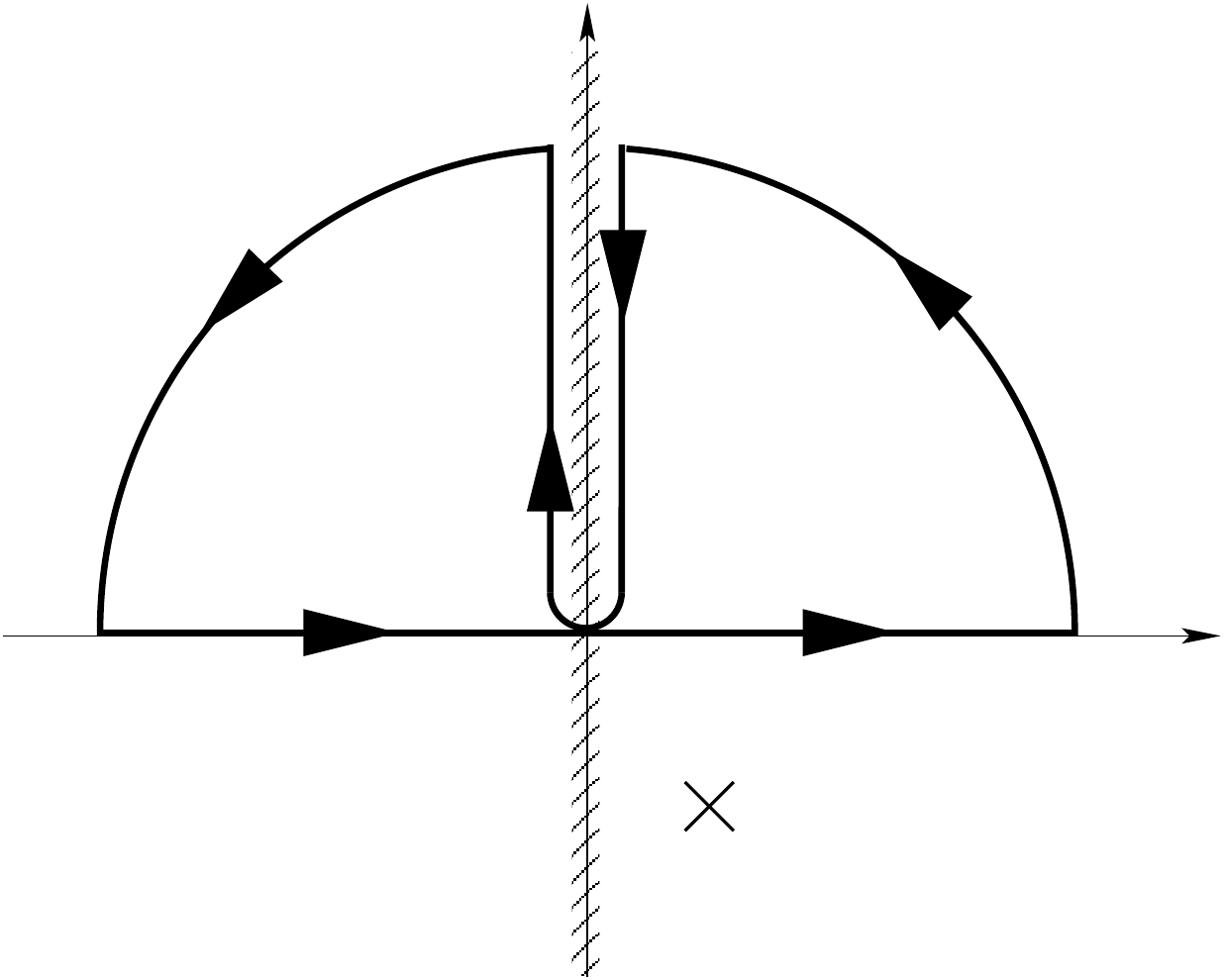}} \caption{Contour of integration : the cross stands for a pole and the hatched line for a branch cut.} \label{contour} \end{figure}

A logarthmic singularity in $q_x$ arises when we integrate over $q_y$, and keeping only this singular part, we write
 \[ \Pi_{a}^{(1)} \sim  \frac{\bar g \rho_{0}\alpha^\prime}{8\pi^2}T\sum_{m \neq 0} \frac{i\alpha^\prime \Omega_{m}}{\gamma |\Omega_{m}|}\int d q_x \ \frac{q_x^2 \ {\rm Log}  ( \Lambda / |q_x | ) }{ ( i\Omega_m   - \alpha^\prime v_F q_x )^3 }  \ . \]  $\Pi_{a}^{(1)}$ is performed by continuation in the upper half plane, if $\Omega_m \leq 0 $ and in the lower half plane if $\Omega_m \geq 0 $ so that to avoid the pole in the Green's function (See Fig. \ref{contour}). 

Changing variables in $q_x= i z$ we get
 \bea
 &\Pi_{a}^{(1)} & =  - \frac{\bar g \rho_{0}\alpha^\prime}{8\pi^2}T\sum_{m \neq 0} \frac{i\alpha^\prime \Omega_{m}}{\gamma |\Omega_{m}|} \times \nonumber \\
&&\int_0^\Lambda   i d z    {\rm Sgn }( \Omega_m ) 
 \frac{(- i z)^2 \left ( {\rm Log } ( -i z) - {\rm Log } ( i z) \right ) }{ (-i)^3  \left (|\Omega_m  |  + \alpha^\prime v_F z \right )^3 }\ ;  \nonumber   \\
 & & =   - \frac{\bar g \rho_{0}\alpha^\prime}{8\pi^2}T\sum_{m \neq 0} \frac{i\alpha^\prime \Omega_{m}}{\gamma |\Omega_{m}|} \times \nonumber \\
&&\frac{ i \pi {\rm Sgn} ( \Omega_m)}{ ( \alpha^\prime v_F)^3} \ {\rm Log} \left ( \frac{\Lambda}{| \Omega_m   | } \right )  \  . \nonumber \eea 

To perform the summation over $m$, we notice that 
 $T \sum_{- \Lambda /T }^{\Lambda/T} 1  = 2 \Lambda $ is independent of T. The same sum without the $m=0$ term will be $2 \Lambda - T$ and to logarithmic accuracy, we obtain : 
 $$T\sum_{m \neq 0} {\rm Log} \left ( \frac{\Lambda}{| \Omega_m   | } \right ) = - T {\rm Log} \left ( \frac{ \Lambda}{T} \right ) + ...$$
 where the dots stand for $O(T)$ terms.
 
 Finally
 \beq \Pi_a^{(1)} (T)  =   - \frac{V^2}{8 \alpha^\prime D^2}   T {\rm Log} \left ( \frac{ \Lambda}{T} \right )  \ . \eeq    
 
 This term is of negative sign and dominant compared to $E^*$, thus it can destabilize the regime. It therefore puts the intermediate regime in a fragile situation.  This is due to the presence of the fermion loop $g_F$. 
  \begin{figure}[h]{\includegraphics[width=8cm]{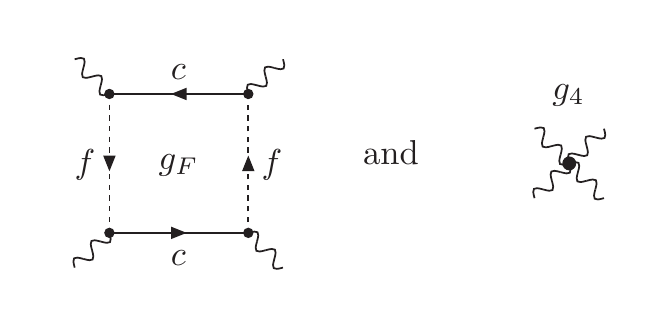}}\caption{The fermionic loop and the $g_4$ vertex} \end{figure}
  \newline
  
 Fortunately, in $D=2$, a mode-mode coupling constant $g_4$, coming for exemple from the term $-J n_i n_j /4$ in Eqn.(\ref{hamiltonian}), provides corrections to scaling of the same temperature dependance but with a positive signe $ - T log T$, competing with the one calculated previously. Indeed, for $g_4 \geq 0$, the $\phi^4$-theory is stable and, close to a QCP, the corrections to scaling follow the law\cite{zinn}
 \begin{figure}
[h]{\includegraphics[width=6cm]{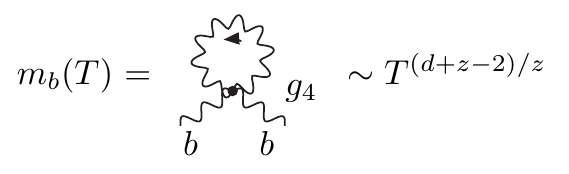}}
 \end{figure}
 
 Precisely, the leading logarithmic contribution coming from the $g_4$ vertex reads
 \bea 
 m_b(T) & = &\frac{g_4 D_0}{6 \pi} T {\rm Log} \left ( \frac{1}{T} \right ) \nonumber \\ 
 & = & g_4\frac{2 D}{3 V^2}T {\rm Log} \left ( \frac{1}{T} \right )
 \eea

 The stability of the intermediate regime is then a matter of prefactors between the two terms. It can lie on a fragile basis, as it requires strong enough ferromagnetic short range fluctuations. However, it has been shown that this regime is stable for $D=3$\cite{cathlong}, we can thus expect that a small three-dimensional character could cure this instability. The correction to the boson mass coming from the gauge fluctuation
 \[  \parbox[c]{3cm}{\includegraphics[width=3cm]{diag3}} \ .  \]
  goes like $T^2$ in $D=2$ and are sub-dominant \cite{cathlong}.
 
  \section{Expression for the free energy}
\label{app:FE}
We start with the Luttinger-Ward formula (\ref{LW}). $F_{MF}$ is the sum of the free energies of the two bands, given by:
$$F_{MF} = -  T\sum_{\pm}2 \,\,\pi \rho_{\pm}\sum_{n}|\omega_{n}|.$$
The sum over the fermionic Matsubara frequencies is formally divergent but its temperature dependence can be extracted using the following spectral representation
$$|\omega_{n}| = -\frac{1}{\pi}\int\frac{x dx}{x-i\omega_{n}},$$ then, performing the summation over Matsubara frequencies, with
$$T \sum_{n}\frac{1}{x-i\omega_{n}} =  \frac{1}{2}-n_{F}(\omega_{n}),$$
we get
$$T\sum_{n} |\omega_{n}| \rightarrow \frac{\pi T^2}{6}$$

We turn now to the bosonic part of (\ref{LW})
\bea
F_{h} & = &\frac{T}{2}\sum_{m}\int\frac{dq^2}{(2\pi)^2}\log{\left [D^{-1}(\mathbf{q},\Omega_{m})\right ]} \nonumber \\
& = & \frac{T}{4\pi}\sum_{m}\int_{0}^{+\infty} q dq \log{\left [q^2 + \xi^{-2} + \frac{\gamma |\Omega_{m}|}{\alpha^\prime q}\right ]} \nonumber
\eea
The integral over the holon momentum is dominated by large momenta, and we have
\bea
F_{h} & \approx & \frac{T}{4\pi}\sum_{m} \int^{+\infty}_{0} \frac{\gamma |\Omega_{m}|}{\alpha^\prime q (q^2 + \xi^{-2})}\nonumber \\
& = & \frac{\gamma \xi}{8\alpha^\prime}T\sum_{m}|\Omega_{m}| \nonumber
\eea
Summation over the bosonic Matsubara frequencies is performed in the same way as for the sum over fermionic frequencies, and we find, for the T-dependent part of it
$$T\sum_{m} |\Omega_{m}| \rightarrow - \frac{\pi T^2}{3}$$
We end up with the total free energy given by
\beq   
 F = - \frac {\pi T^2}{6} \left [ 2 \pi ( \rho_+ + \rho_- ) + \frac{ \gamma \xi}{ 4 \alpha^\prime} \right ] \ . \eeq

\end{document}